# ANISOTROPIES IN THE COSMIC MICROWAVE BACKGROUND: AN ANALYTIC APPROACH [†]

Wayne Hu[1] and Naoshi Sugiyama[1,2]

[1]*Departments of Astronomy and Physics*
*University of California, Berkeley, California 94720*

[2]*Department of Physics, Faculty of Science*
*The University of Tokyo, Tokyo, 113, Japan*

We introduce a simple yet powerful *analytic* method which obtains the structure of cosmic microwave background anisotropies to better than 5-10% in temperature fluctuations on *all* scales. It is applicable to *any* model in which the potential fluctuations at recombination are both linear and known. Moreover, it recovers and explains the presence of the "Doppler peaks" at degree scales as *driven* acoustic oscillations of the photon-baryon fluid. We treat in detail such subtleties as the time dependence of the gravitational driving force, anisotropic stress from the neutrino quadrupole, and damping during the recombination process, again all from an analytic standpoint. We apply this formalism to the standard cold dark matter model to gain physical insight into the anisotropies, including the dependence of the peak locations and heights on cosmological parameters such as $\Omega_b$ and $h$, as well as model parameters such as the ionization history. Damping due to the finite thickness of the last scattering surface and photon diffusion are further more shown to be identical. In addition to being a powerful probe into the nature of anisotropies, this treatment can be used in place of the standard Boltzmann code where 5-10% accuracy in temperature fluctuations is satisfactory and/or speed is essential. Equally importantly, it can be used as a portable standard by which numerical codes can be tested and compared.

*Subject Headings:* Cosmology: Cosmic Microwave Background; Cosmology, Theory

hu@pac1.berkeley.edu, sugiyama@pac2.berkeley.edu

*It is the nature of things that they* are *ties to each other.*
–*Chuang-tzu*

## 1. Introduction

After their discovery by the COBE satellite (Smoot *et al.* 1992), cosmic microwave background (CMB) anisotropies have become one of the most powerful observational probes of cosmology. Indeed recent detections of CMB anisotropies on degree scales (see *e.g.* White, Scott, & Silk 1994 and references therein) provide us with important information about large scale structure formation in the universe. It is consequently of great interest to understand their origin. To predict CMB anisotropies in any given model, we have to solve the coupled equations for the evolution of all species present. Namely, these are the Euler and continuity equations for the fluid components (the baryons and cold dark matter) and the Boltzmann equations for the massless components (the photons and neutrinos). These coupled equations have been numerically solved by many authors (*e.g.* Peebles & Yu 1970, Wilson & Silk 1981, Bond & Efstathiou 1984, Vittorio & Silk 1984). Although it is sufficient for direct comparison of a specific model with observations, this "black box" approach makes it difficult to extract the physical content of the predictions. This problem is compounded by the fact anisotropy formation is a rather complicated process. It has long been known that several physically distinct effects contribute to their generation, *e.g.* the gravitational redshift (Sachs & Wolfe 1967), the adiabatic growth of perturbations (Peebles & Yu 1970), baryon velocity induced fluctuations (Zel'dovich & Sunyaev 1970), and photon diffusion (Silk 1968). An analytic treatment is therefore desirable to gain physical insight into CMB anisotropies.

In this *paper*, we present a *fully* analytic treatment for the evolution of CMB temperature perturbations and systematically investigate each contribution to the final observable anisotropy. It can be applied to *any* model with the standard thermal history, regardless of dark matter content or initial conditions. Several analytic calculations of CMB anisotropies have been performed in the past under less general, and often unrealistic, assumptions. Doroshkevich, Zel'dovich, & Sunyaev (1978) and Doroshkevich (1988) presented an analytic expression for temperature anisotropies on the last scattering surface. Based on this work, Naselsky and Novikov (1993), Jørgensen *et al.* (1994), and Atrio-Barandela & Doroshkevich (1994) have recently calculated the anisotropies in the cold dark matter (CDM) scenario. However these works did not realistically account for the evolution of the gravitational potential inside the Jeans scale during radiation domination or on any scale through matter-radiation equality. Indeed CMB anisotropies are quite sensitive to this evolution as we shall show. Thus agreement with the numerical solutions could not be established in those works. Furthermore, they used an overly simplistic account of fluctuation evolution during the recombination process when the damping scale of CMB anisotropies is fixed. They also neglected the neutrino contributions to the anisotropic stress which is important during radiation domination.

Moreover the synchronous gauge condition employed by most previous treatments makes the separation and physical interpretation of these effects difficult. Here we use gauge invariant perturbation theory (Bardeen 1980; Kodama & Sasaki 1984) where each physical process is readily distinguished. Recently a gauge invariant treatment of CMB anisotropies has been performed by Seljak (1994) based on similar approximations to our own. However, it employs a numerical solution to the tightly coupled evolution equations. Our analytic treatment allows one to separate and individually examine each contribution to the anisotropy easily, yet still maintains sufficient generality to realistically describe a model such as CDM.

In §2, we introduce the general technique and present the central results of our analytic approximation, valid for any gravitational instability model. Starting from the exact multifluid differential equations (§2.1), we derive the tight coupling approximation (§2.2) and obtain its analytic solution in terms of simple quadratures (§2.3). The tight coupling approximation is modified to include photon diffusion through recombination in §2.4. All of these approximations are treated in greater detail in Appendix B. Furthermore, useful analytic formulae which describe the recombination process are presented in Appendix C. In §2.5, we discuss the free streaming solution and show how the final observable anisotropies are obtained.

We specialize these techniques to the CDM model in §3. The so-called "Doppler peaks" in the anisotropy spectrum are shown to arise from driven acoustic oscillations in the tight coupling regime. The analytic form of the gravitational driving force is derived in Appendix A and summarized in §3.1, including effects due to radiation pressure and anisotropic stress. We show that the location (§3.2) and the heights (§3.3) of the peaks as well as all other features in the anisotropy can be simply understood from a physical standpoint in this formalism. Moreover the predictions of our analytic treatment are accurate to better than the $5-10\%$ level in temperature perturbations on *all* scales, when compared with the full numerical solutions based on Sugiyama & Gouda (1992). Finally in Appendix D, we provide a step by step recipe for constructing the analytic solutions.

## 2. The Evolution of Perturbations

An exact solution for the evolution of the cosmological perturbations involves solving the coupled evolution equations for all of the species present. It is quite apparent that this can only be accomplished through numerical integration. However, components such as the neutrinos and the cold dark matter are only coupled to the photons and baryons gravitationally, whereas before recombination the photons and baryons are tightly coupled by Compton scattering. It is thus sufficient to consider the simpler problem of tracking the evolution of a single, tightly coupled, baryon-photon fluid in a gravitational potential that accounts for the other species. As we shall now show, this problem naturally lends itself to analytic solution for it can be described as an oscillator whose restoring force is given by the photon pressure and whose driving term is determined by the gravitational potentials.

*2.1 General Equations*

The evolution equation for the $k$th Fourier mode of the gauge invariant Newtonian temperature perturbation* $\Theta(\eta,\mu)$ is given by the Boltzmann equation with a source from Compton scattering,

$$\dot{\Theta} + ik\mu(\Theta + \Psi) = -\dot{\Phi} + \dot{\tau}[\Theta_0 - \Theta - \frac{1}{10}\Theta_2 P_2(\mu) - i\mu V_b], \tag{1}$$

where overdots are derivatives with respect to the conformal time $\eta = \int dt(a_0/a)$, $k\mu = \mathbf{k}\cdot\boldsymbol{\gamma}$, with $\gamma_i$ as the direction cosines of the photon momentum, and $\dot{\tau} = x_e n_e \sigma_T a/a_0$ is the differential optical depth to Thomson scattering. Here $x_e$ is the ionization fraction, $n_e$ is the total electron density, $\sigma_T$ is the Thomson scattering cross section, $c=1$, and the scale factor $a/a_0 = (1+z)^{-1}$. The gauge invariant metric perturbations are $\Psi$ the Newtonian potential, and $\Phi$ the perturbation to the intrinsic spatial curvature. We will refer to both $\Psi$ and $\Phi$ as "gravitational potentials." These terms are simply related to the total density fluctuations through the generalized Poisson equation and the anisotropic stress (see Appendix A). Note that $\Phi = -\Psi$ when anisotropic stress is negligible, *e.g.* in the matter dominated limit. In equation (1), we employ conventions where the multipole decomposition is given by

$$\Theta(\eta,\mu) = \sum(-i)^\ell \Theta_\ell(\eta) P_\ell(\mu), \tag{2}$$

---

\* For brevity, when discussing a *single k* mode of the perturbation, we drop the implicit $k$ index of the variables. For example, $\Theta(\eta,\mu)$ should be understood as $\Theta(\eta,\mu,k)$. Where confusion may arise, *e.g.* in the discussion of initial and final power spectra, we restore it. Real space fluctuations do not appear in this paper.

and have chosen the amplitude of the baryon velocity $\gamma \cdot \mathbf{v}_b = -i\mu V_b$ accordingly. The appearance of the photon quadrupole $\Theta_2$ in equation (1) merely represents the angular dependence of Compton scattering. Finally, we have also assumed a flat geometry. For the open universe generalization of all the arguments presented here see Hu & Sugiyama (1994b). A less technical summary of those results may be found in Hu (1994).

Notice that the gravitational potentials $\Psi$ and $\Phi$ have two effects on the temperature fluctuations both introduced in the original Sachs & Wolfe (1967) paper. The gradient of the Newtonian potential $\Psi$ induces a gravitational redshift on the photons as they travel through the potential well. Since the potential difference merely induces a fractional temperature shift of the same magnitude, the combination $\Theta + \Psi$ is the resultant temperature perturbation after the photon climbs out of a well of negative $\Psi$. We will consequently often use $\Theta + \Psi$ to describe the effective perturbation rather than $\Theta$ alone. This accounts for what we call the *ordinary* Sachs-Wolfe effect. The time dependence of the metric term $\Phi$ causes its own time-dilation effect referred to here as the *integrated* Sachs-Wolfe (ISW) effect.

On the other hand, the baryons evolve under the continuity and Euler equations

$$\dot{\Delta}_b = -k(V_b - \Theta_1) + \frac{3}{4}\dot{\Delta}_\gamma \,,$$
$$\dot{V}_b = -\frac{\dot{a}}{a}V_b + k\Psi + \dot{\tau}(\Theta_1 - V_b)/R \,, \qquad (3)$$

where $R = 3\rho_b/4\rho_\gamma$ is the scale factor normalized to $3/4$ at *photon-baryon* equality. Here $\Delta_b$ and $\Delta_\gamma$ are the baryon and photon energy density perturbations in the total matter rest frame representation (see Appendix A). Note that the Newtonian potential acts as a source to the velocity through infall and gives rise to the adiabatic growth of perturbations. This coupled photon-baryon system, described by equations (1) and (3), fully determines the CMB anisotropies. Notice that the *only* effect of the other decoupled components is through the potentials $\Psi$ and $\Phi$.

## 2.2 Tight Coupling Limit

Before recombination, the differential optical depth $\dot{\tau}$ is high making Compton scattering extremely rapid and effective. Together equations (1) and (3) then imply that $V_b = \Theta_1$ and $\Theta_\ell = 0$, for $\ell \geq 2$. This merely reflects the fact that scattering makes the photon distribution isotropic in the electron rest frame. Equation (3) then tells us that $\dot{\Delta}_b = 3/4\dot{\Delta}_\gamma$, *i.e.* the evolution is *adiabatic*. Correspondingly, the resultant temperature fluctuations are also adiabatic. This should be distinguished from temperature fluctuations in reionized scenarios which are generated by Doppler shifts off electrons at last scattering. In that case, a relatively unperturbed photon distribution with $V_b \gg \Theta_1$ receives a Doppler shift from the last scattering event. Here the photons are *already* isotropic in the electron rest frame, *i.e.* $V_b = \Theta_1$, implying *no* shift at last scattering: the photons merely decouple at recombination. Only where tight coupling breaks down, *i.e.* below the diffusion length, can a Doppler effect arise. However since the term "Doppler peak" is so firmly entrenched in the literature, we will continue to use it to describe anisotropies from the acoustic oscillations.

The tight coupling approximation involves expanding the Boltzmann and Euler equations in the Compton scattering time $\dot{\tau}^{-1}$ to eliminate the baryonic variables (Peebles & Yu 1970). To first order, we obtain a *single* second order differential equation (see Appendix B),

$$\ddot{\Theta}_0 + \frac{\dot{a}}{a}\frac{R}{1+R}\dot{\Theta}_0 + k^2 c_s^2 \Theta_0 = F(\eta) \,, \qquad (4)$$

where the forcing function $F(\eta)$ arises from the gravitational potentials and is given by

$$F(\eta) = -\ddot{\Phi} - \frac{\dot{a}}{a}\frac{R}{1+R}\dot{\Phi} - \frac{k^2}{3}\Psi \,. \qquad (5)$$

Here the photon-baryon sound speed $c_s$ is

$$c_s^2 = \frac{1}{3}\frac{1}{1+R} \qquad (6)$$

from which we obtain the sound horizon,

$$r_s(\eta) = \int_0^\eta c_s d\eta' \,. \qquad (7)$$

Equation (4) tells us that, aside from expansion damping, there are three major tight coupling evolutionary effects with different spheres of influence:

[1] $\ddot{\Phi}$, the ISW effect on $\ddot{\Theta}_0$ which, when present, dominates at superhorizon scales $k\eta \ll 1$;

[2] $k^2\Psi$, the gravitational infall, which leads to the adiabatic growth of the photon-baryon fluctuations and becomes important near the horizon scale $k\eta \sim 1$;

[3] $k^2 c_s^2 \Theta_0$, the photon pressure which cannot be neglected inside the sound horizon $kr_s = k\int c_s d\eta \gtrsim 1$.

The ordinary Sachs-Wolfe effect, which occurs when the photons stream out of the perturbation *after* tight coupling breaks down, can be taken into account with the combination $\Theta + \Psi$. This partially cancels the infall term since it counters the gravitational shift experienced by the photon as it falls into the gravitational well.

The infall and pressure terms must of course compete since the latter prevents the former from causing adiabatic growth at sufficiently small scales. The scale at which these forces are in balance is known as the Jeans scale. Below this scale, equation (4) describes acoustic oscillations of the photon-baryon fluid. By including both the gravitational driving force *and* the pressure in the intermediate regime crucial for degree scale anisotropies, we thus refine the simple Jeans instability argument.

*2.3 Solutions in the Tight Coupling Limit*

Solutions in the tight coupling approximation are straightforward to write down and easily manipulated into useful forms (see Appendices B & D). Equation (4) is simply that of an forced, damped oscillator. Therefore, the homogeneous $F(\eta) = 0$ equation can be solved by the WKB method, in the limit where the frequency is slowly varying. Naturally, the solutions are oscillatory functions with phase $\phi = kr_s = k\int c_s d\eta$. The WKB approximation is valid for all modes that are in the oscillating regime by last scattering, *i.e.* smaller than the sound horizon at recombination. The particular solution, denoted with an overhat $\hat{\Theta}_0$, may be found from Green's method,

$$[1 + R(\eta)]^{1/4}\hat{\Theta}_0(\eta) = \Theta_0(0)\cos kr_s(\eta) + \frac{\sqrt{3}}{k}[\dot{\Theta}_0(0) + \frac{1}{4}\dot{R}(0)\Theta_0(0)]\sin kr_s(\eta) \\ + \frac{\sqrt{3}}{k}\int_0^\eta d\eta'[1+R(\eta')]^{3/4}\sin[kr_s(\eta) - kr_s(\eta')]F(\eta') \,. \qquad (8)$$

Furthermore, the dipole solution can be obtained from the zeroth moment of equation (1), *i.e.* the photon continuity equation $k\Theta_1 = -3(\dot{\Theta}_0 + \dot{\Phi})$. An examination of equation (8) shows that this implies the dipole oscillates $\pi/2$ out of phase with the monopole and has a factor $\dot{r}_s = c_s \propto (1+R)^{-1/2}$ suppression in amplitude.

The evolution of modes larger than the sound horizon at last scattering can be obtained by taking $R = 0$ in equation (8) as proven in Appendix B. The two solutions can be simply joined at $k_s = 0.08h^3$ Mpc$^{-1}$ if last scattering occurs sufficiently before photon-baryon equality $R(\eta_*) \ll 1$, as is the case for the standard big bang nucleosynthesis scenarios with $\Omega_b \ll \Omega_0$.

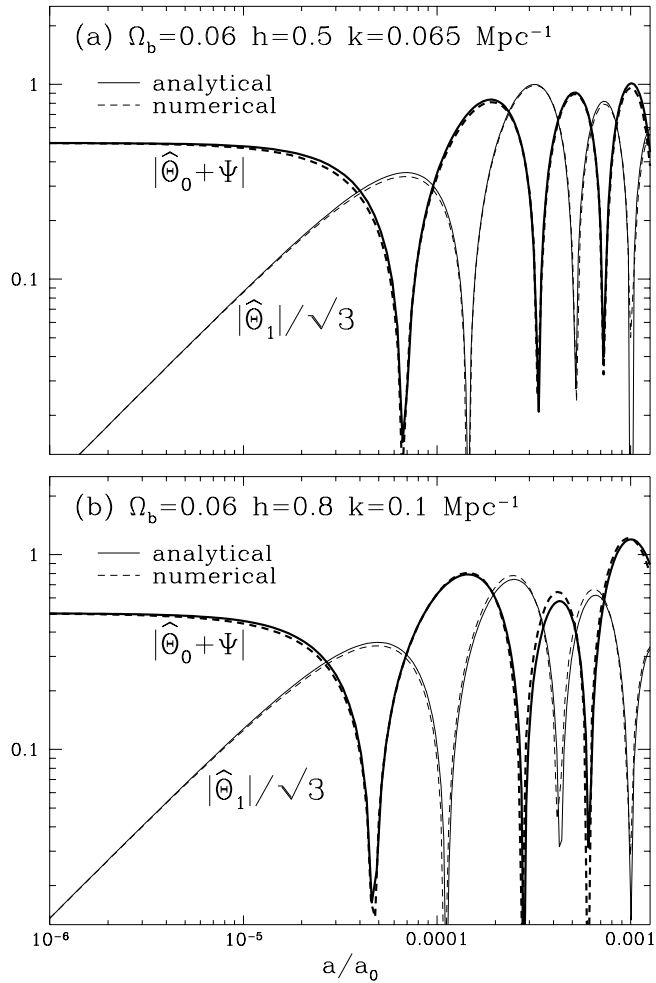

**Figure 1.** Temperature fluctuation evolution before recombination. The tight coupling approximation obtains the evolution to high accuracy compared with the full numerical solution once the potentials are known. Here two CDM models are taken as examples. Notice that in the high $h$ case, the dipole $\hat{\Theta}_1$ is significantly suppressed with respect to the monopole $\hat{\Theta}_0 + \Psi$, and the monopole oscillations themselves are severely modulated. The $\sqrt{3}$ in the amplitude of the dipole accounts for its three degrees of freedom. Here and here alone, the numerical results are for a universe which never recombines to eliminate diffusion damping at recombination. The arbitrary overall normalization has been set here and in the following two figures to $\Psi(0) = -1$.

Already we can gain useful insight on the structure of anisotropies. Scales which reach an extrema in the monopole at last scattering will yield a corresponding peak in the anisotropy power spectrum. Note that both the positive and negative extrema in temperature yield peaks in the power spectrum. The zeros of the monopole will be partially filled in by the dipole but still represent troughs in the final anisotropy pattern. As an example, we display in Fig. 1 the tight coupling evolution in the CDM model for two choices of the Hubble constant $H_0 = 100h$ km s$^{-1}$ Mpc$^{-1}$: (a) h=0.5 and (b) h=0.8. We have also plotted the numerical solution to the full set of perturbation equations for comparison. The small discrepancy is in fact almost entirely due to slight inaccuracies in the analytic form of the CDM potentials from Appendix A (see also Fig. 3). Notice that the suppression of the dipole is more effective in the high $h$ case since $R \propto \Omega_b h^2$. Furthermore the amplitude of the monopole oscillations are severely modulated in this high $h$ case. This feature, due to gravitational enhancement of the compressional phase and suppression of the expansion phase, is further discussed in §3.3.

In summary, once the potentials $\Phi$ and $\Psi$ are known, the tight coupling solution for the temperature perturbation can be readily evaluated with equation (8) at any desired epoch before recombination. In the large scale and oscillatory regime, the analytic approximation is limited only by the accuracy with which we know the potentials.

*2.4 Diffusion Damping and Recombination*

The first order tight coupling solutions presented in §2.3 can only apply on scales much larger than the mean free path of the photons. This scale grows with the expansion and becomes essentially infinite through recombination. To handle this photon diffusion problem, we can expand equations (1) and (3) to second order in the Compton scattering time $\dot{\tau}^{-1}$. The well known result [see *e.g.* Peebles (1980) and Appendix B] is that the tight coupling solution $\hat{\Theta}_0$ [equation (8)] is exponentially damped,

$$(\Theta_0 + \Psi) = (\hat{\Theta}_0 + \Psi)e^{-[k/k_D(\eta)]^2}, \qquad (9)$$

where the diffusion scale is essentially the distance a photon can random walk by $\eta$,

$$k_D^{-2}(\eta) = \frac{1}{6} \int_0^\eta d\eta \frac{1}{\dot{\tau}} \frac{R^2 + 4(1+R)/5}{(1+R)^2}. \qquad (10)$$

We give the analytic form of $\dot{\tau}(\eta)$, *i.e.* the ionization fraction $x_e(\eta)$, valid through recombination in Appendix C.

This severe damping described by equation (9) arises because as the photons diffuse through the baryons, fluctuations become anisotropies which are exponentially damped by scattering (Silk 1968). Here the presence of $\Psi$ represents the ordinary Sachs-Wolfe contribution as mentioned in §2.1. As photons diffuse in and out of potential wells, they continue to pick up gravitational redshifts. These contributions are only dependent on potential differences and are not damped out by diffusion. Moreover, this treatment automatically accounts for the increase in the diffusion length at recombination. As the ionization fraction $x_e(\eta) \to 0$, $\dot{\tau} \to 0$ and the diffusion length becomes infinite. Of course, most photons by definition last scatter before the diffusion length tends toward infinity. This implies that fluctuations are severely damped under the "thickness" of the last scattering surface, *i.e.* the average diffusion length for a photon at last scattering. Previous work on the tight coupling approximation, *e.g.* Jørgensen *et al.* (1994), Atrio-Barandela & Doroshkevich (1994), and Seljak (1994), have all treated the effects of diffusion and recombination damping separately in a rather *ad hoc* manner.

The structure of the "Doppler peaks" can be *completely* described by these damped, driven adiabatic oscillations in any scenario where last scattering is sufficiently early. For reionized universes, last scattering is delayed, and the diffusion length grows to be nearly the horizon at last scattering. In this case, degree scale anisotropies are no longer determined by the adiabatic effect since $V_b \gg \Theta_1$ but rather by Doppler shifts as photons diffuse across the baryons at last scattering. Analytic techniques for studying this situation are well known and also accurate at the 10% level in temperature fluctuations (see *e.g.* Hu, Scott, & Silk 1994; Hu & Sugiyama 1994a, Dodelson & Jubas 1994 and references therein). These diffusive techniques should *not* be used to describe standard recombination scenarios.

## 2.5 The Free Streaming Solution

Aside from the diffusion modification contained in equation (9), the adiabatic fluctuations present at recombination are merely frozen in and free stream to the present (Bond & Efstathiou 1987). They will consequently be observable as anisotropies in the microwave background sky today. Let us formalize this statement. Equation (1) has the solution (Hu & Sugiyama 1994a)

$$[\Theta + \Psi](\eta_0, \mu) = \int_0^{\eta_0} \left\{ [\Theta_0 + \Psi - i\mu V_b]\dot{\tau} - \dot{\Phi} + \dot{\Psi} \right\} e^{-\tau(\eta,\eta_0)} e^{ik\mu(\eta-\eta_0)} d\eta \,, \tag{11}$$

where the optical depth is measured from $\eta$ to the present epoch $\eta_0$, $\tau(\eta_1, \eta_2) = \int_{\eta_1}^{\eta_2} \dot{\tau} d\eta$, and we have dropped the quadrupole term since it vanishes in the tight coupling limit. The combination $\dot{\tau} e^{-\tau}$ is called the conformal time visibility function and is the probability that a photon last scattered within $d\eta$ of $\eta$. Naturally it has a sharp peak at the last scattering epoch $\eta_*$. For improvements on the Jones & Wyse (1985) fitting formulae for this epoch and recombination in general, see Appendix C.

Taking the multipole moments and setting $V_b = \Theta_1$, we find for $\ell \geq 2$,

$$\begin{aligned}
\Theta_\ell(\eta_0) &\approx [\Theta_0 + \Psi](\eta_*)(2\ell + 1) j_\ell(k\Delta\eta_*) \\
&\quad + \Theta_1(\eta_*)[\ell j_{\ell-1}(k\Delta\eta_*) - (\ell+1) j_{\ell+1}(k\Delta\eta_*)] \\
&\quad + (2\ell + 1) \int_{\eta_*}^{\eta_0} [\dot{\Psi} - \dot{\Phi}] j_\ell(k\Delta\eta) d\eta \,,
\end{aligned} \tag{12}$$

where $\Delta\eta = \eta_0 - \eta$, $\Delta\eta_* = \eta_0 - \eta_*$. The fluctuations on the last scattering surface are determined from the *undamped* WKB solution (8) as $[\Theta_0 + \Psi](\eta_*) = [\hat{\Theta}_0 + \Psi](\eta_*) \mathcal{D}(k)$ and $\Theta_1(\eta_*) = \hat{\Theta}_1(\eta_*) \mathcal{D}(k)$, where the $k$ index of the perturbations is again suppressed, and we have employed equation (9) to obtain the average damping factor,

$$\mathcal{D}(k) = \int_0^{\eta_0} \dot{\tau} e^{-\tau(\eta,\eta_0)} e^{-[k/k_D(\eta)]^2} d\eta. \tag{13}$$

Here we have also assumed that all functions save the damping factor are slowly varying compared to the visibility function. This is a good approximation on all scales of interest but breaks down for extremely small scales where both the oscillations in $j_\ell$ and $\hat{\Theta}_0$ are rapid. If the potentials are not exactly constant after $\eta_*$, as is the case if matter-radiation equality occurs close to or after $\eta_*$, the integral in equation (12) yields an ISW effect *after* last scattering. Notice that the potential today at the observer does not contribute to anisotropies so that the ordinary Sachs-Wolfe effect is given entirely by $\Theta + \Psi$ at last scattering. We do not include the present dipole since it cannot be separated from the peculiar motion of the observer.

Since $j_\ell(x)$ has a peak at $\ell \sim x$, equation (12) merely represents the free streaming conversion of a perturbation on a spatial scale at last scattering to an angular scale on the sky today. An example of these last scattering surface fluctuations, we show the analytic results for the CDM scenario in Fig. 2. It is important to note that the phase information between the monopole and dipole displayed here is preserved in the free streaming transformation. It is *not* sufficient to free stream the rms sum of the monopole and dipole (Bond & Efstathiou 1987) if we want to obtain the detailed structure of the peaks. In particular, equation (12) tells us that power in the dipole is distributed more broadly in $\ell$ than the monopole.

Integrating over all $k$ modes of the perturbation, we obtain

$$\frac{2\ell+1}{4\pi} C_\ell = \frac{V}{2\pi^2} \int \frac{dk}{k} \frac{k^3 |\Theta_\ell(\eta_0, k)|^2}{2\ell+1}, \tag{14}$$

where we have restored the $k$ index implicit in equation (12), and $C_\ell$ is normalized in the standard manner such that the observed $(\Delta T/T)_{rms}^2 = \sum_\ell (2\ell+1) C_\ell W_\ell / 4\pi$ with $W_\ell$ as the experimental window function. We now have the full analytic apparatus to calculate CMB anisotropies in any given model where the gravitational potential and recombination history are specified.

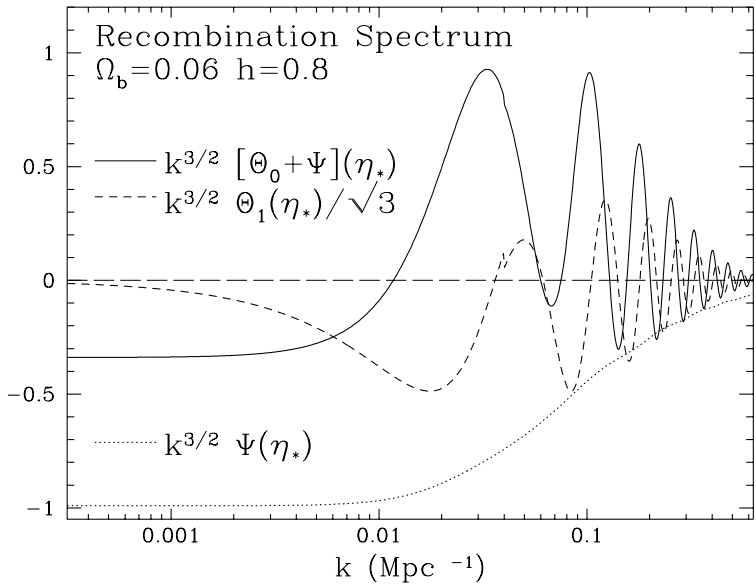

**Figure 2.** Analytic temperature fluctuation spectrum at recombination. Again a Harrison-Zel'dovich CDM model is chosen as an example. Fluctuations on the last scattering surface free stream to the observer creating anisotropies. The phase relation between the monopole and dipole as well as their relative amplitudes give rise to the structure of the Doppler peaks. Notice that the dipole is significantly smaller than the monopole as expected but is not negligible, especially near the zeros of the monopole oscillations. In particular, along with the ISW effect, it fills in fluctuations *before* the first Doppler peak. At intermediate scales, gravity is able to shift the equilibrium position of the fluctuations leading to a modulation of the monopole peaks (see §3.2). We have drawn in the zero level of the oscillations to guide the eye. The kink at $k = 0.04$ Mpc$^{-1}$ is due to the joining of the large and small scale solutions. At large scales, the Sachs-Wolfe effect dominates, bringing the effective anisotropy to $[\Theta_0 + \Psi](\eta_*) \approx \frac{1}{3}\Psi(\eta_*)$.

## 3. Anisotropies in the CDM Model

The formalism developed in §2 is applicable to any set of initial conditions or decoupled dark matter and may even be readily generalized to cosmological constant or open models. In fact, the open isocurvature case is treated in Hu & Sugiyama (1994b). We shall show in this section that it is a powerful and accurate technique for calculating and understanding CMB anisotropies by focusing on the standard CDM model. This scenario has decoupled cold dark matter supplying the dark mass to make $\Omega_0 = 1$ as well as the usual photon, baryon and massless neutrino components. In addition, inflation predicts the initial spectrum of total density fluctuations to be nearly Harrison-Zel'dovich and adiabatic in the initial conditions (see §A.3 for the specific definition). For definiteness, we will present the results of this adiabatic Harrison-Zel'dovich spectrum. Let us now consider the CMB anisotropies in this CDM model.

*3.1 Gravitational Potentials and Large Scale Anisotropies*

In the spatially flat $\Omega_0 = 1$ CDM model we consider here, the potentials $\Phi$ and $\Psi$ are constant outside the Jeans scale in the radiation dominated epoch and on all scales in the matter dominated epoch. However it is important to note that they *decay* between Jeans crossing in the radiation dominated epoch and full matter domination (see Fig. 3). Furthermore, $\Psi = -\Phi$ only in the matter dominated limit when anisotropic stress becomes negligible. The constant potential approximation, used as a toy model by Jørgensen *et al.*

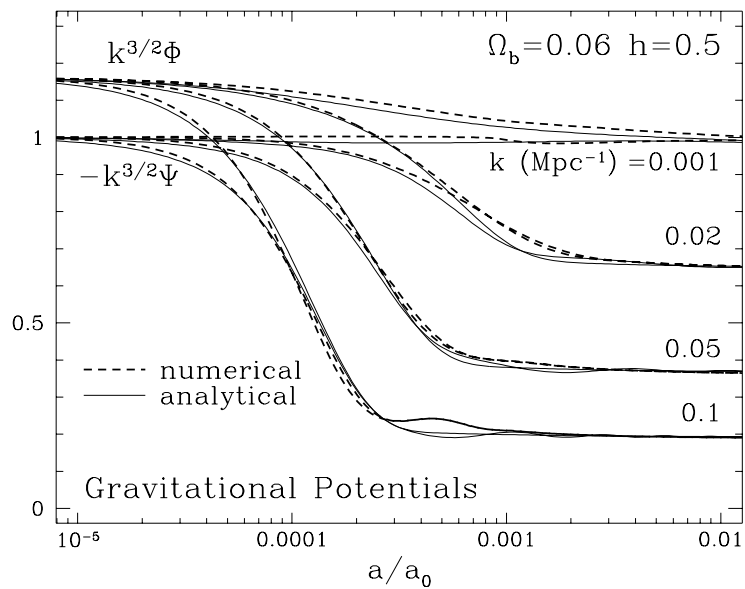

**Figure 3.** Gravitational potentials in the Harrison-Zel'dovich CDM model. Notice that the potential decays after crossing the Jeans scale in the radiation dominated epoch and only flattens out *well* into the matter dominated epoch. Moreover $\Phi \neq -\Psi$ early on due to anisotropic stress. These two facts have significant consequences for the temperature anisotropies. Notice that the analytic approximations of Appendix A trace the numerical potentials reasonably well.

(1994) and Atrio-Barandela & Doroskevich (1994), is *not* a good description of the CDM scenario. The qualitative difference is that in the CDM scenario, the driving term of the oscillations becomes ineffective once the Jeans scale grows much larger than the perturbation size. This leads to a more prominent first "Doppler peak" and the more complex structure of higher peaks in the real spectrum.

To understand large scale anisotropies, let us first take the simple zeroth order approximation by neglecting anisotropic stress and taking the universe to be completely matter dominated at last scattering. In this case, the analysis of Appendix A tells us that $\Theta_0(\eta_*) \approx -2/3\Psi(\eta_*) \approx -2/3\Psi(\eta_0)$. The sign accounts for the fact that fluctuations are larger deep in a potential well. The magnitude is given by the initial conditions and the ISW effect. Climbing out of the potential $\Psi(\eta_*)$ after last scattering, the photons are left with a final anisotropy given by $[\Theta_0 + \Psi](\eta_*) \approx 1/3\Psi(\eta_0)$ as is well known.

However Fig. 3 shows us that anisotropic stress cannot be ignored in the radiation dominated limit since $\Psi(0) \neq -\Phi(0)$, and the potentials are not precisely constant until well into the matter dominated epoch. The former problem is treated in detail in Appendix A by solving perturbatively for the anisotropic stress. The time variation of the potential on the other hand leads to an ISW effect *after* last scattering and must be included if better than 10% accuracy in temperature fluctuations on scales up to and including the first Doppler peak is required. This is especially important for low $h$ models where matter-radiation equality occurs near recombination as mentioned in §2.5. Because it is an *integrated* effect, and the angle subtended by a given scale increases as time progresses, it tends to spread anisotropies among the low $\ell$'s making the rise to the Doppler peaks more gradual than it would otherwise be (see Appendix A.4). This effect and the presence of the dipole peak *before* the first Doppler peak (see §3.2) explains why anisotropies do not behave as $\ell(\ell + 1)C_\ell =$ constant for Harrison-Zel'dovich initial conditions as the ordinary Sachs-Wolfe effect would imply (Abbott & Schaefer 1986).

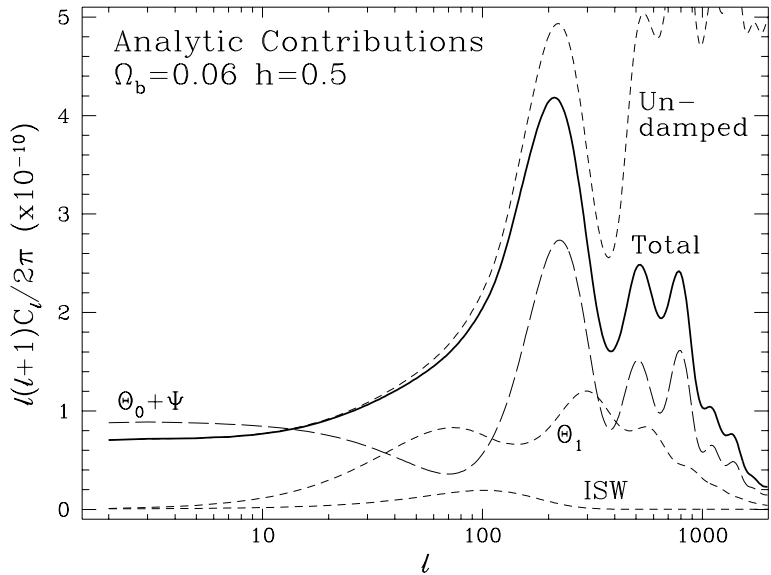

**Figure 4.** Individual contributions to the anisotropy in the Harrison-Zel'dovich CDM model. At the largest scales ($\ell \lesssim 30$), the monopole $|\Theta_0 + \Psi|$ from the ordinary Sachs-Wolfe effect dominates. The 20% correction from the *post*-recombination ISW effect on scales larger than the first Doppler peak appears misleadingly small in power (see text). The ordinary Sachs-Wolfe effect is overpowered by adiabatic growth of the monopole at small scales leading to a deficit at intermediate scales ($\ell \sim 70$) which is filled in by the adiabatic dipole $\Theta_1$ and the ISW effect. Although the dipole cannot be neglected, the monopole is clearly responsible for the general structure of the Doppler peaks. Diffusion damping significantly reduces fluctuations beyond the first Doppler peak and cuts off the anisotropies at $\ell \sim 1000$.

*3.2 Mapping the Anisotropy Spectrum*

Now let us consider the general features of the anisotropies and determine the scales at which each physical process dominates. In Fig. 4, we separate and individually examine the various contributions to the anisotropy. In both the large and small scale limits, the monopole $\Theta_0 + \Psi$ at recombination governs the structure of the anisotropies. Scales much larger than the angle subtended by the sound horizon at recombination are dominated by the ordinary Sachs-Wolfe effect. However, the correction from the *post*-recombination ISW effect represented in Fig. 4 appears misleadingly small in power. Note that the 20% shift in power spectrum normalization from the monopole-only solution is entirely due to the 1% ISW effect. This is explained by the fact that the ISW effect adds nearly coherently with the monopole whereas the dipole roughly adds in quadrature. Because most of the contribution to the ISW effect occurs near matter-radiation equality where $\eta_0 - \eta \approx \eta_0$, crudely speaking the ISW integral in equation (12) can be estimated as

$$\int_{\eta_*}^{\eta_0} [\dot{\Psi} - \dot{\Phi}] j_\ell(k\Delta\eta) d\eta \approx [\Psi - \Phi]\Big|_{\eta_*}^{\eta_0} j_\ell(k\eta_0). \tag{15}$$

The ISW effect therefore mimics the ordinary Sachs-Wolfe effect. In fact, the it partially cancels the ordinary Sachs-Wolfe effect at large scales since the potential wells become *shallower* as time progresses. Indeed, this effect contributes at the 20% level in power on all scales up to and including the first Doppler peak for low $h$ models. The extent to which equation (15) is a good approximation is discussed in Appendix A.4.

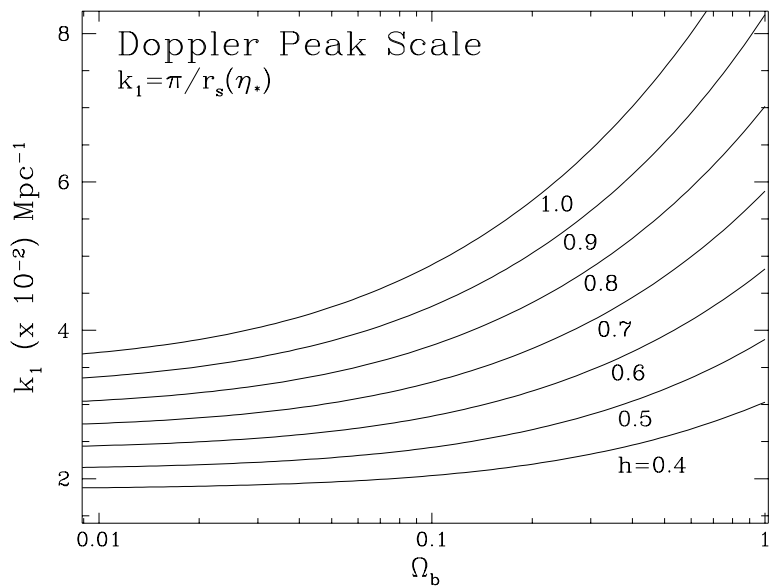

**Figure 5.** Location of the first Doppler peak. The Doppler peaks are determined by the extrema of the monopole oscillations at last scattering $k_m = m\pi/r_s(\eta_*)$. Because the sound horizon becomes independent of $\Omega_b$ as $\Omega_b h^2$ goes to zero, the location of the peak is nearly independent of $\Omega_b$ if $\Omega_b \ll \Omega_0$. The integral over conformal time in $r_s$ makes $k_m \propto mh$ so that the angular location of the peaks $\ell_m \approx k_m \eta_0 \approx m k_1 \eta_0$ is nearly independent of $h$ as well.

Near the horizon at recombination, the adiabatic growth of the photon energy density due to infall into the potential wells is more than sufficient to make up for the redshift from climbing out. Under the sound horizon, photon pressure makes these adiabatic perturbations oscillate and leads to the structure of the Doppler peaks.

The full analytic solution given in equation (8) recovers the location of the peaks to excellent accuracy. However, it may be useful to break this solution down to show the intuitive origin of this result. As an examination of equation (8) shows, any solution to the tight coupling equations is of the form

$$\Theta_0(\eta) = A_1(\eta) \cos k r_s(\eta) + A_2(\eta) \sin k r_s(\eta). \tag{16}$$

For adiabatic fluctuations, gravitational infall contributes primarily to the $\cos k r_s$ harmonic. This is because the potential is constant until Jeans scale crossing at which point it begins to decay. The driving term therefore mimics $\cos k r_s$ and causes the monopole to follow suit. This also implies that the dipole goes as $\sin k r_s$ and vanishes at scales much larger than the sound horizon $k r_s \ll 1$. In contrast, isocurvature fluctuations, which begin with vanishing potential, have monopoles which are dominated by the $\sin k r_s$ mode as one would expect. These tendencies hold exactly in the small scale limit (Hu & Sugiyama 1994b). Yet even for the first few peaks, it is a reasonable approximation to take $\cos k r_s$ as the dominant adiabatic solution.

Therefore the peaks in the temperature power spectrum will be located at the scale $k_m$ which satisfy $k_m r_s(\eta_*) = m\pi$ where $m$ is an integer $\geq 1$. Note that this locates the first Doppler peak at roughly the sound horizon which is close to, but conceptually distinct from the Jeans scale. The troughs at $k_{m-1/2} r_s(\eta_*) = (m - 1/2)\pi$ will be partially filled in by the dipole, including $k_{1/2}$ which occurs *before* the first Doppler peak (Stompor 1994, see also Fig. 4). In Fig. 5, we plot the scale $k_1$ corresponding to the location of the first

Doppler peak as a function of $\Omega_b$ for various choices of $h$. Since $kr_s(\eta) = k \int c_s \eta = (k/\sqrt{3}) \int (1+R)^{-1/2} d\eta$, its dependence on $\Omega_b h^2$ is weak in the CDM model where $\Omega_0 \gg \Omega_b$ and $R(\eta_*) \ll 1$. Furthermore, $c_s$ becomes independent of $h$ implying that $r_s \propto \eta$. Therefore the peak modes are $k_m \propto mh$.

Now let us consider the $\ell$ space structure of the peaks. Since the last scattering surface is located at approximately the horizon distance $\eta_0$, the angle subtended by the scale $k_m$ will roughly correspond to $\ell_m \approx k_m \eta_0 \approx mk_1 \eta_0$, and the oscillatory structure of the fluctuations will be partially preserved. Notice that the peaks in the dipole are more washed out than the monopole due to the broader nature of the $k$ to $\ell$ conversion in equation (12) for the dipole. Furthermore, since $\eta \propto h^{-1}$, the $h$ dependence of the $\ell$ space peaks $\ell_m$ cancels. For the low value of $\Omega_b$, which is required by nucleosynthesis, the location of the peaks will consequently be roughly independent of both $\Omega_b$ and $h$ as is well known. Note however that due to gravitational suppression of the even peaks (see §3.3) often *only* the odd $m$ peaks are distinguishable.

The last significant feature in the spectrum is the diffusion cut off. In Appendix C, we show that the optical depth $\tau$ near recombination is nearly independent of $h$ and only weakly dependent on $\Omega_b$. Therefore from equation (10), the damping scale becomes $k_D(\eta_*) \propto h$ at recombination. Since the location of the peaks scale as $k_m \propto h$, the cutoff in $\ell$ is essentially independent of $h$ and very mildly dependent on $\Omega_b$. The effect of this damping is displayed in Fig. 4. Employing the undamped solutions $\hat{\Theta}_0$ and $\hat{\Theta}_1$ causes a gross overestimate of the small scale anisotropies. The problem is compounded in $\ell$ space since each $\ell$ mode in reality gets contributions from a range of $k$ modes, including $k > k_D$. See Appendix C for further discussion including the separation of the damping at recombination from that which occurs before it.

*3.3 Heights of the Peaks*

To understand the heights of the Doppler peaks, a somewhat more intricate argument is necessary. Let us begin with the first Doppler peak. There are essentially two effects that govern its behavior. Firstly, the gravitational infall $\propto \Psi$ competes with the restoring force from the pressure $\propto c_s^2 \Theta_0$. This would naively imply that the deeper the potential, the larger the resultant fluctuation. As we show in Appendix A.4, increasing $h$ makes matter-radiation equality earlier and the potential at the peak deeper due to less pressure damping. However anisotropies do *not* increase with $h$ for *all* $\Omega_b$. A deep potential implies a large redshift as the photons climb out after last scattering by the ordinary Sachs-Wolfe effect. Moreover if the sound horizon is approximately the particle horizon $r_s = \int c_s d\eta \sim \eta$, the window in which adiabatic growth dominates is small (see §2.2). Thus, a deeper potential implies a *smaller* final anisotropy if the pressure is held fixed at a high value, *i.e.* $c_s \sim 1/\sqrt{3}$.

Therefore the way to increase the anisotropies at the first peak is to reduce the pressure, *i.e.* the sound speed, rather than deepen the potential. This can be accomplished if we raise $\Omega_b h^2$. The height of the first Doppler peak thus *increases* with $\Omega_b$ (Holtzman 1989, Fukugita, Sugiyama, Umemura 1990). As for the $h$ dependence, we have uncovered two opposing effects. Since the sound speed becomes $c_s \approx 1/\sqrt{3}$, independent of $h$ as $\Omega_b$ goes to zero, the variation of the pressure with $h$ asymptotically disappears. This is exactly the limit where the Sachs-Wolfe mechanism is most effective at counterbalancing adiabatic growth. Therefore at low $\Omega_b$ raising $h$ will *decrease* the anisotropy whereas at high $\Omega_b$ it will *increase* the anisotropy. The crossover point happens to coincide approximately with the nucleosynthesis value of $\Omega_b \approx 0.05$ if $h \approx 0.5 - 0.8$. Thus the $h$ dependence of the *first* Doppler peak for standard CDM will be relatively weak. Note however that the Sachs-Wolfe mechanism becomes less important at smaller scales due to the decay of the potential, whereas the pressure argument is scale free.

Now let us consider the general structure of the peaks. For all odd numbered peaks, the situation will be qualitatively identical to that of the first Doppler peak: $\Psi$ and $\Theta_0$ have opposite signs corresponding to overdensities in potential wells and underdensities at the peaks. In other words, inside the gravitational well, the oscillation is in its compressional phase and is thus *enhanced* by gravitational infall. On the other hand, the even numbered peaks correspond to expansion inside the well and are correspondingly suppressed. The more gravity dominates over pressure, *i.e.* the higher $\Omega_b h^2$ is, the more effective is this pattern of alternating enhancements and suppressions. In some cases, the second Doppler peak disappears altogether. This modulation effect can also be thought of as a shift in the equilibrium point of the oscillations due to the gravitational force (*c.f.* Fig. 2). Thus it disappears at high $k$ where the potential becomes negligible. Since in the high $\Omega_b h^2$ case the dipole is also severely suppressed with respect to the monopole peaks, this implies that in these models the decrement between the first few positive (odd) peaks will be substantial.

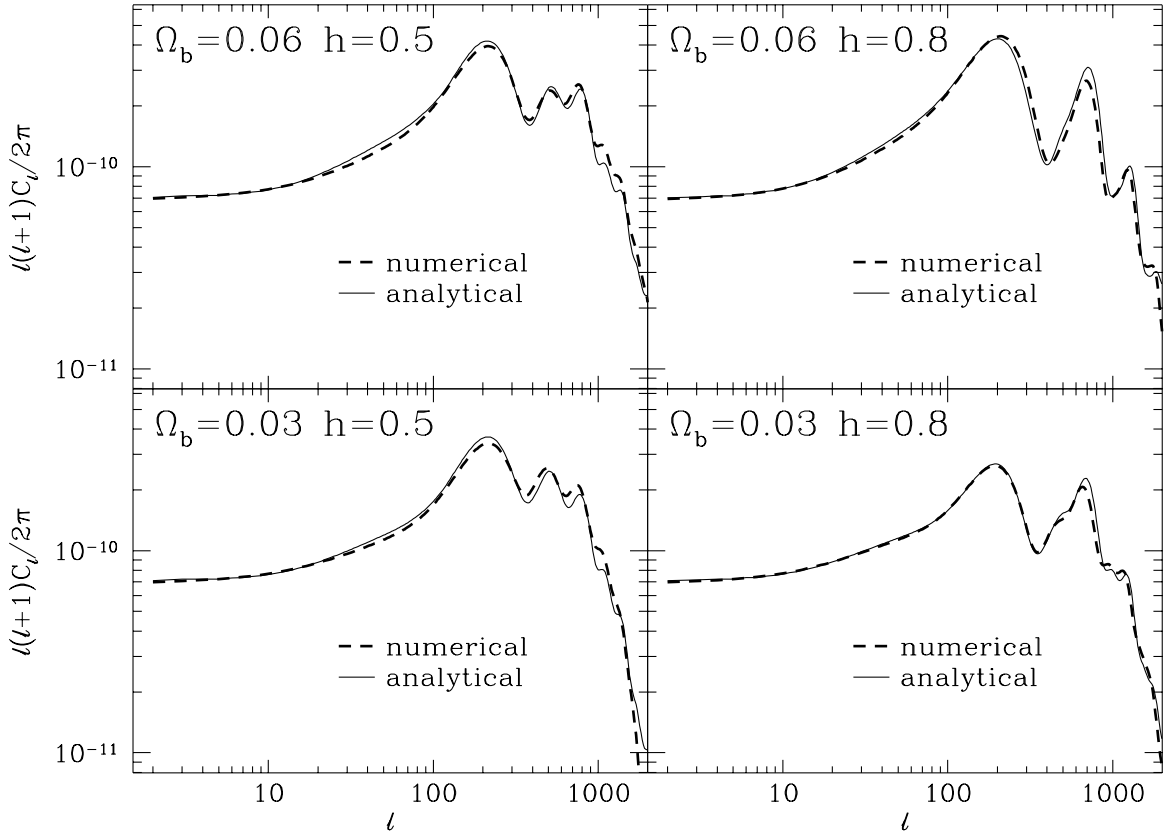

**Figure 6.** The CDM anisotropy spectrum. Notice that for high $\Omega_b$ raising $h$ increases the amplitude of the peaks, due to a lowering of the pressure, whereas for low $\Omega_b$ raising $h$ decreases it, due to a deepening of the potential well out of which the photon must climb. The pressure argument also explains the monotonic increase in the heights with $\Omega_b$. The high $\Omega_b h^2$ case also has a significant decriment between the first and third Doppler peaks due to the modulation effect and a significantly smaller dipole contribution.

In Fig. 6, we display the final anisotropy spectrum $C_\ell$ for various choices of $\Omega_b$ and $h$. Notice that for $\Omega_b = 0.06$, raising $h$ increases the amplitudes of the peaks, whereas for $\Omega_b = 0.03$ it decreases the amplitudes of the peaks. Furthermore, the modulation effect described above shows up for the high $\Omega_b h^2$ case making the even peaks invisible. In all cases, the analytic solution matches the full numerical results typically to better than 10% in power. Our analysis therefore includes *all* the major contributions to the anisotropy.

## 4. Conclusions

We have presented a fully analytic treatment of CMB anisotropies under the assumption of tight coupling between the baryons and photons before recombination. Our technique describes the tightly coupled photon-baryon fluid as an oscillator in an arbitrary potential well. Since dark matter affects this system only through these potential wells, *all* dark matter models can be treated under this formalism. Applying this method to the CDM model, we show that it typically obtains CMB anisotropies to 10% or better in power on all scales. Moreover it allows us to separate and explain each and every process that contributes to the final anisotropy. On the very large scale, the familiar Sachs-Wolfe tail dominates the CMB spectrum. However the flatness of this tail is broken by both the adiabatic photon dipole contribution *and* the ISW term, particularly for small $h$ models because the universe is not totally matter dominated at last scattering surface. The series of "Doppler peaks" results from extrema in the oscillatory monopole at last scattering. These acoustic oscillations are driven by the gravitational potentials which cause the *adiabatic* growth of density fluctuations. In fact the name "Doppler peaks" is itself somewhat misleading since it implies that they result from Doppler shifts induced by last scattering off electrons in infall. Since anisotropies in a reionized scenario *are* generated by such a mechanism, it is perhaps preferable to call these oscillatory fluctuations "adiabatic peaks" instead.

We moreover obtain the location and heights of these peaks to excellent accuracy and show that they are easily explained by variations in the sound speed and the depth of the potential wells. Thus we have also clarified the dependence of CMB anisotropies on cosmological parameters, *i.e.* the Hubble constant $h$ and the baryon fraction $\Omega_b$. Furthermore, by taking the recombination process into account more properly than previous work (see Appendix C), we recover the damping scale from photon diffusion arguments alone. This implies that the damping due to the finite thickness of last scattering surface and photon diffusion are the one and the same.

Our interpretation of these physical processes is confirmed by its excellent agreement with the full numerical solutions of the coupled evolution equations. Readers can even use our method, reconstructed in detail in Appendix D, instead of full Boltzmann code if only 10% accuracy in power is required. In fact, our formalism is useful in combination with numerical work as well. It can be employed as a simple, portable standard with which various numerical treatments can be compared. Moreover, in the large scale and oscillatory limit, the approximations presented here are limited only to the extent that the gravitational potentials are known. Since the evolution of density fluctuations can be solved far more easily than the infinite heirarchy of temperature multipoles, our formalism can be used as a quick and simple way to accurately calculate and understand the CMB anisotropy in *any* model where the fluctuations are still linear at recombination.

## Acknowledgements

We would like to thank E. Bunn, D. Scott, J. Silk, and M. White for useful discussions. W.H. has been partially supported by the NSF. N.S. acknowledges financial support from a JSPS postdoctoral fellowship for research abroad.

# References

Abbott & Schaefer 1986, ApJ, 308, 546.

Atrio-Barandela, F. & Doroshkevich, A.G. 1994, ApJ, 420, 26.

Bardeen, J.M. 1980, Phys. Rev., D22, 1882.

Bardeen, J.M., Bond, J.R., Kaiser, N. & Szalay, A.S. 1986, ApJ, 304, 15.

Bennett, C.L. *et al.* 1994, ApJ, (in press)

Bond, J.R. & Efstathiou, G. 1984, ApJ, 285, L45.

Bond, J.R. & Efstathiou, G. 1987, MNRAS, 226, 665.

Dodelson, S. & Jubas, J, 1994, ApJ (in press)

Doroshkevich, A.G. 1988, Sov. Astron. Lett., 14, 125.

Doroshkevich, A.G., Zel'dovich, Ya. B., & Sunyaev, R.A. 1978, Sov. Astron., 22, 523.

Fukugita, M., Sugiyama, N., & Umemura, M. 1990, 358, 28.

Gorski, K., *et al.* 1994, ApJ, (in press).

Holtzman, J.A. 1989, ApJ Supp., 71, 1

Hu, W. 1994, in CWRU CMB Workshop: 2 Years after COBE, eds. L. Krauss & P. Kernan, (World Scientific, Singapore, in press).

Hu, W., Scott D., & Silk, J. 1994, Phys. Rev. D, 49, 648.

Hu, W. & Sugiyama, N. 1994a, Phys. Rev. D, (in press).

Hu, W. & Sugiyama, N. 1994b, Phys. Rev. D (to be submitted).

Jones, B.J.T & Wyse, R.F.G. 1985, Astron. Astrophys., 149, 144.

Jørgensen, H.E., Kotok, E., Naselsky, P., & Novikov, I. 1994, Astron. Astrophys., (in press).

Kodama, H. & Sasaki, M. 1984, Prog. Theor. Phys. Suppl., 88, 1.

Kodama, H. & Sasaki, M. 1986, Int. J. Mod. Phys., A1, 265.

Kodama, H. & Sasaki, M. 1987, Int. J. Mod. Phys., A2, 491.

Naselsky, P. & Novikov, I. 1993, ApJ, 413, 14.

Peacock, J.A. & Dodds, S.J. 1994, MNRAS, 267, 1020.

Peebles, P.J.E. 1968, ApJ, 153, 1.

Peebles, P.J.E. & Yu, J.T. 1970, ApJ, 162, 815.

Peebles, P.J.E. 1980, The Large Scale Structure of the Universe, (Princeton University Press, Princeton, NJ).

Sachs, R.K. & Wolfe, A.M. 1967, ApJ, 147, 73.

Seljak, U. 1994, ApJ Lett, (submitted).

Silk, J. 1968, ApJ, 151, 459.

Stompor, R. 1994, Astron. Astrophys., (in press)

Sugiyama, N. & Gouda, N. 1992, Prog. Theor. Phys., 88, 803.

Smoot, G., *et al.* 1992, ApJ, 396, L1.

Sunyaev, R.A. & Zel'dovich, Ya.B. 1970, A.S.S., 7, 3.

Vittorio, N. & Silk, J. 1984, ApJ, 285, 39.

Wilson, M.L. & Silk, J. 1981, ApJ, 243, 14.

White, M., Scott, D. & Silk, J. 1994, Ann. Rev. Astro. Astrophys., (in press).

## Appendix A: Gravitational Potentials and Anisotropic Stress

*A.1 General Relations*

The potential $\Phi$ is related to the total density fluctuation $\Delta_T = \delta\rho_T/\rho_T$ by the generalized Poisson equation

$$k^2\Phi = 4\pi G\rho(a/a_0)^2\Delta_T, \tag{A-1}$$

and thus accounts for the decoupled cold dark matter and neutrinos. Throughout this appendix, we normalize the scale factor $a$ at matter-radiation equality. In order to determine the potential, we must analyze the contributions from all species. In fact, the evolution equations for the CDM and the massless neutrinos can be obtained from equations (1) and (3) by the replacements $\Delta_b \to \Delta_c$ and $\Theta \to N$ with $\dot{\tau} \to 0$ where $\Delta_c$ and $N$ are the density fluctuation in the CDM and the temperature perturbation of the neutrinos respectively. These replacements are valid since the only difference between these species in the non-degenerate limit is Compton scattering.

We also need the Newtonian potential $\Psi$ which is related to $\Phi$ through the anisotropic stress $\Pi$,

$$\Phi + \Psi = -\frac{8\pi G p}{k^2}\left(\frac{a}{a_0}\right)^2\Pi, \tag{A-2}$$

where $\Pi$ is given by the quadrupole moments of the the photons and neutrinos,

$$p\Pi = \frac{12}{5}(p_\gamma\Theta_2 + p_\nu N_2). \tag{A-3}$$

If either the pressure $p$ is unimportant or the quadrupole moments vanish, $\Psi = -\Phi$. We know that $\Theta_2$ vanishes due to the isotropizing effect of Compton scattering in the tight coupling limit. Even $N_2$ can only be generated through free streaming from $N_0$. Our approach to obtaining an analytic solution for $\Phi$ and $\Psi$ is therefore to include $\Pi$ as a small perturbation.

Combining the four evolution equations, we obtain (Kodama & Sasaki 1984)

$$\begin{aligned}\dot{\Delta}_T - 3w\frac{\dot{a}}{a}\Delta_T &= -(1+w)kV_T - 2\frac{\dot{a}}{a}w\Pi, \\ \dot{V}_T + \frac{\dot{a}}{a}V_T &= \frac{4}{3}\frac{w}{(1+w)^2}k\Delta_T + k\Psi - \frac{2}{3}k\frac{w}{1+w}\Pi,\end{aligned} \tag{A-4}$$

where $w = p/\rho$, and the total velocity is given by the sum over component velocities $(\rho+p)V_T = \sum_i(\rho_i+p_i)V_i$, with $V_\gamma \equiv \Theta_1$ and $V_\nu \equiv N_1$. Here we have assumed adiabatic conditions and have also used the relations

$$\begin{aligned}\Delta_\gamma &= 4\Theta_0 + 4\frac{\dot{a}}{a}\frac{V_T}{k}, \\ \Delta_\nu &= 4N_0 + 4\frac{\dot{a}}{a}\frac{V_T}{k},\end{aligned} \tag{A-5}$$

in converting from the Newtonian variables to the total matter rest frame variables.

*A.2 Zeroth Order Large Scale Solution*

In the zeroth order approximation where $\Pi = 0$, equation (A-4) has exact solutions in the limit that we can ignore the $k\Delta_T$ pressure term in the velocity equation. This is appropriate for superhorizon sized fluctuations.* The growing mode $\bar{\Delta}_T(a) = AU_G(a)$ is given by (Kodama & Sasaki 1986),

$$U_G(a) = \left[ a^3 + \frac{2}{9}a^2 - \frac{8}{9}a - \frac{16}{9} + \frac{16}{9}\sqrt{a+1} \right] \frac{1}{a(a+1)}, \tag{A-6}$$

where recall that the scale factor is normalized at matter-radiation equality $a_{eq} = 1$. Here the normalization factor $A(k)$ defines the relative weighting of the $k$ modes through the initial power spectrum (see A.3). On the other hand for reference, the decaying mode takes the form

$$U_D(a) = \frac{1}{a\sqrt{a+1}}. \tag{A-7}$$

Note that $U_G = (10/9)a^2$ in the radiation dominated limit and $a$ in the matter dominated epoch. This gives the well known result that outside the horizon, the potential is constant in both the radiation and matter dominated limits. In this $\Pi = 0$ limit, we obtain $\bar{\Psi}(\eta_0) = 9/10\bar{\Psi}(0)$. Equation (A-6) with (A-4) and (A-5) also implies that $\bar{\Theta}_0(0) = -1/2\bar{\Psi}(0)$.

This behavior for the potentials of course has significant consequences for CMB anisotropies. Equation (1) tells us that at large scales where the ISW effect dominates, $\dot{\bar{\Theta}}_0 = -\dot{\bar{\Phi}}$. This implies that the fluctuations at last scattering, which determine the *ordinary* Sachs-Wolfe effect, are given by

$$\bar{\Theta}_0(\eta_*) = \bar{\Theta}_0(0) + \bar{\Psi}(\eta_*) - \bar{\Psi}(0), \tag{A-8}$$

since $\bar{\Phi} = -\bar{\Psi}$. If last scattering occurs well into the matter dominated epoch, the potential is constant and $\bar{\Psi}(\eta_*) = \bar{\Psi}(\eta_0)$. Putting these results together, we obtain $\bar{\Theta}_0(\eta_*) = -2/3\bar{\Psi}(\eta_*)$ and the total Sachs-Wolfe temperature perturbation after the photons climb out of the potential,

$$[\bar{\Theta}_0 + \bar{\Psi}](\eta_*) = \frac{1}{3}\bar{\Psi}(\eta_0), \tag{A-9}$$

which is the familiar Sachs & Wolfe (1967) result. Indeed, this zeroth order solution can in fact be employed at large scales if only 10% accuracy in the temperature fluctuations even at the COBE DMR normalization scale is acceptable. Of course, an error at the normalization scale causes an error on all scales. It is therefore preferable to correct for the small effect of anisotropic stress perturbatively.

---

* Since inside the Jeans scale, pressure acts to damp the growth of perturbations, we will denote the *un*damped large scale solution with an overbar analogous to our notation for the tight coupling solutions. This is to caution the reader that the solutions only apply prior to horizon crossing for all modes that cross before matter-radiation equality.

*A.3 First Order Large Scale Solution*

Before recombination, due to the isotropizing effects of scattering, the anisotropic stress of the photons is negligible small. Hence the main contribution to $\Pi$ comes from the neutrino quadrupole anisotropy $N_2$ [see (A–3)]. Here we analytically obtain the growing mode solution for density perturbations including the contribution of anisotropic stress for modes outside the horizon at matter-radiation equality. This together with the matter transfer function is sufficient to obtain the gravitational potential on all scales.

We take into account the neutrino quadrupole anisotropy perturbatively. Namely, we use the exact zeroth order solutions (A–6) and (A–7) to obtain the anisotropic stress. We then take this solution to iteratively correct for anisotropic stress in equation (A–4). If we neglect higher order multipole components, which is reasonable for superhorizon sized modes, the second moment of the the Boltzmann equation for the neutrino becomes

$$\dot{N}_2 = \frac{2}{3}kN_1 \approx \frac{2}{3}kV_T , \tag{A-10}$$

where we employ causality to infer that the fluid cannot separate at superhorizon scales, *i.e.* $N_1 \equiv V_\nu \approx V_T$ (see Hu & Sugiyama 1994b). Moreover the exact zeroth order solution for $V_T$ is obtained using equation (A–6) in the first of equations (A–4), yielding the solution to equation (A–10),

$$\bar{N}_2(a) \approx 2A \int_0^a \frac{da'}{a'} \frac{1}{3a'+4} \left( U_G - (a'+1)a'\frac{dU_G}{da'} \right) , \tag{A-11}$$

where we have used the relation $3w = 1/(1+a)$ and recall that the overbar represents the *undamped* superhorizon solution. Although it is possible to analytically integrate equation (A–11), the expression is cumbersome. Instead, we can employ an approximate solution which is exact in the limit $a \ll 1$ and $a \gg 1$,

$$\bar{N}_2(a) = -\frac{1}{10}\frac{20a+19}{3a+4}AU_G - \frac{8}{3}\frac{a}{3a+4}A + \frac{8}{9}\ln\left(\frac{3a+4}{4}\right)A . \tag{A-12}$$

We have checked that this approximation works quite well by comparing it to equation (A–11) and the full numerical solution.

Next, we employ the above solution for $\bar{N}_2$ in equations (A–4). These two first order equations may be rewritten as one second order equation for $\Delta_T$. Analogously to the tight coupling equation (4), the particular solution including the source terms $\Pi$ and $\dot{\Pi}$ can be obtained from the homogeneous solutions $U_G$ and $U_D$ by Green's method,

$$\bar{\Delta}_T(a) = \left(1 + \frac{2}{5}f_\nu\right)AU_G(a) + \frac{2}{5}f_\nu[I_1(a)U_G(a) + I_2(a)U_D(a)] , \tag{A-13}$$

where $I_1(a) = \int_0^a da' S(a')U_D(a')$, $I_2(a) = \int_0^a da' S(a')U_G(a')$,

$$S(a) = \frac{24}{5}\frac{(a+1)^{5/2}}{a^2(3a+4)}\left\{\frac{2a}{3a+4}\frac{d}{da}AU_G(a) \right.$$
$$\left. -\frac{2}{(3a+4)(a+1)}AU_G(a) + \left[\frac{1}{(a+1)^2} - \frac{2}{a+1} + \frac{12}{3a+4}\right]\bar{N}_2(a)\right\} , \tag{A-14}$$

and $f_\nu$ is the ratio of neutrino to total radiation density $f_\nu \equiv \rho_\nu/(\rho_\nu + \rho_\gamma)$. If we assume three massless neutrinos and the standard thermal history, $\rho_\nu/\rho_\gamma = 3(7/4)(4/11)^{4/3}/2 = 0.68$, *i.e.* $f_\nu = 0.405$. The first term in equation (A–13) comes from the initial conditions for $\Delta_T$ which can be iteratively established by employing equation (A–10) in (A–4). All terms which are proportional to $f_\nu$ in the equation (A–13) come from equation (A–3) since the anisotropic stress $\Pi \approx (12/5)f_\nu N_2$. The asymptotic behavior of the equation (A–13) is

$$\bar{\Delta}_T(a) \to \left(1 + \frac{2}{5}f_\nu\right) AU_G(a) \quad (a \ll 1)$$
$$\to \left(1 + \frac{2}{5}f_\nu(1 - 0.333)\right) AU_G(a) \quad (a \gg 1). \quad (A-15)$$

Here we have used the fact that if $a \gg 1$, the decaying term $I_2 U_D$ may be ignored and $I_1 \to -0.333$.

Therefore we may obtain a simple approximate expression for the large scale density fluctuations,

$$\bar{\Delta}_T(a) \approx \left[1 + \frac{2}{5}f_\nu \left(1 - 0.333\frac{a}{a+1}\right)\right] AU_G(a). \quad (A-16)$$

Again we have checked this approximation works reasonably well by comparing it to numerical calculations. The potentials $\bar{\Phi}$ and $\bar{\Psi}$ are therefore written as

$$\bar{\Phi}(a) = \frac{3}{4}\left(\frac{k_{eq}}{k}\right)^2 \frac{a+1}{a^2}\bar{\Delta}_T(a),$$
$$\bar{\Psi}(a) = -\frac{3}{4}\left(\frac{k_{eq}}{k}\right)^2 \frac{a+1}{a^2}\left(\bar{\Delta}_T(a) + \frac{8}{5}f_\nu \frac{\bar{N}_2(a)}{a+1}\right), \quad (A-17)$$

where $k_{eq} = \sqrt{2}(\Omega_0 H_0^2 a_0)^{1/2}$ is the scale that passes the horizon at matter-radiation equality. By using the asymptotic form of $\bar{\Delta}_T$ and $\bar{N}_2$, we easily obtain the corresponding relation between $\bar{\Phi}$ and $\bar{\Psi}$,

$$\bar{\Phi}(a) = -\left(1 + \frac{2}{5}f_\nu\right)\bar{\Psi}(a) \quad (a \ll 1)$$
$$= -\bar{\Psi}(a) \quad (a \gg 1). \quad (A-18)$$

Also of interest are the ratios of initial to final values of the the gravitational potentials: $\bar{\Phi}(a_0) = -\bar{\Psi}(a_0) = 0.86\bar{\Phi}(0)$ and $\bar{\Psi}(a_0) = 1.00\bar{\Psi}(0)$. Thus we see that the correction for anisotropic stress makes a 10% difference in $\bar{\Psi}$ in the radiation dominated epoch. If recombination occurs near equality, this results in a small correction to the standard Sachs-Wolfe formula due to anisotropic stress.

The initial conditions for the perturbations may now be expressed in terms of $\bar{\Phi}(0)$,

$$\Psi(0) \equiv \bar{\Psi}(0) = -0.86\bar{\Phi}(0),$$
$$\Theta(0) \equiv \bar{\Theta}(0) = 0.43\bar{\Phi}(0). \quad (A-19)$$

Note that since all modes are superhorizon sized at the initial epoch, the overbar is superfluous. Moreover, even in the initial conditions, the anisotropic stress represents a small but important correction to the $\Pi = 0$ solutions of §A.2, $\bar{\Phi}(0) = -\bar{\Psi}(0) = 2\bar{\Theta}_0(0)$. Finally, we can relate these quantities to the initial power spectrum,

$$k^3|\Phi(0,k)|^2 \equiv k^3|\bar{\Phi}(0,k)|^2 = \left[\frac{5}{6}\left(1 + \frac{2}{5}f_\nu\right)\right]^2 \left(\frac{k}{k_{eq}}\right)^4 A^2(k) = Bk^{n-1}, \quad (A-20)$$

where we have again restored the implicit $k$ index and the overall normalization factor $B$ is fixed by the COBE DMR detection (see Appendix D). The Harrison-Zel'dovich initial spectrum predicted by inflation is obtained by setting $n = 1$.

*A.4 Small Scale Potentials*

Next we need to obtain solutions of $\Psi$ and $\Phi$ in the small scale limit where pressure cannot be neglected. Qualitatively speaking, we know that the potentials decay inside the Jeans length in the radiation dominated epoch since pressure prevents $\Delta_T$ from growing. However in general, it is impossible to obtain the exact solution valid through matter-radiation equality even if we neglect the anisotropic stress term. Only the asymptotic behaviors in certain limits have been found (Kodama & Sasaki 1987). For the CDM scenario, it is well known that the *final* value of the potential at small scales is obtained from the superhorizon solution (A-17) by the transfer function $\Phi(a_0) = -\Psi(a_0) = T(k)\bar{\Phi}(a_0)$, where

$$T(k) = \frac{\ln(1 + 2.34q)}{2.34q}[1 + 3.89q + (14.1q)^2 + (5.46q)^3 + (6.71q)^4]^{-1/4}, \tag{A-21}$$

with $q \equiv k/[\Omega_0 h^2 \exp(-2\Omega_b)]$ (Peacock & Dodds 1994, Bardeen *et al.* 1986). Note that $q \propto k/k_{eq}$ approximately, reflecting the fact that only modes that cross the Jeans length before equality are suppressed. This implies that the potentials are larger in amplitude if equality occurs later, *i.e.* for high $\Omega_0 h^2$ models. Equation (A-21) therefore empirically accounts for the lack of growth in the radiation dominated era.

Now let us consider the time evolution of the potential. We know that in the matter dominated epoch the potentials are constant on all scales. Therefore, we smoothly join the superhorizon scale solutions of equation (A-17) with a constant matter dominated tail whose relative amplitude is given by the transfer function. Since the Jeans crossing epoch is approximately same as horizon crossing time in radiation dominated era, we can take $(k/Ha) \sim ak/k_{eq} \sim 1$ as the matching epoch,

$$\begin{aligned}\Phi(a) &= \bar{\Phi}(a)\left\{[1-T(k)]\exp[-\alpha_1 (ak/k_{\rm eq})^\beta] + T(k)\right\}, \\ \Psi(a) &= \bar{\Psi}(a)\left\{[1-T(k)]\exp[-\alpha_2 (ak/k_{\rm eq})^\beta] + T(k)\right\},\end{aligned} \tag{A-22}$$

where $\alpha_1$, $\alpha_2$ and $\beta$ are fitting parameters. We also need a small correction to take into account the free streaming oscillations of the neutrino quadrupole inside the Jeans scale. A very simple approximation can be obtained by making the replacement $\bar{N}_2(a) \to \bar{N}_2(a)\cos[0.5k/(Ha)]$ in equation (A-17) for $\Psi(a)$. Here 0.5 is a fitting factor and the Hubble parameter $H(a) = (\dot{a}/a)(a_0/a)$. This crude approximation is sufficient for our purposes. Comparing this functional form (A-22) with numerical results, we obtain a good fit for $\alpha_1 = 0.11$, $\alpha_2 = 0.097$ and $\beta = 1.6$.

In order to calculate the post-recombination ISW effect, we take the direct derivative of equations (A-22). Although this makes the estimation of the derivatives much worse than for the potentials themselves, it is a reasonable approximation in the context of temperature fluctuations since the ISW effect is effectively only a perturbation to the spectrum. Another way of seeing this is to compare the full calculation to the approximation to the ISW integral,

$$\int_{\eta_*}^{\eta_o} [\dot{\Psi} - \dot{\Phi}]j_\ell(k\Delta\eta)d\eta \approx [\Psi - \Phi]\big|_{\eta_*}^{\eta_o} j_\ell(k\eta_0). \tag{A-23}$$

This simple approximation eliminates the need for detailed knowledge of the derivatives by assuming that most the contribution comes from early enough epochs that $j_\ell(k\Delta\eta) \approx j_\ell(k\eta_0)$. Since there is only a contribution near matter-radiation equality, the integral does indeed get most its contribution at early times.

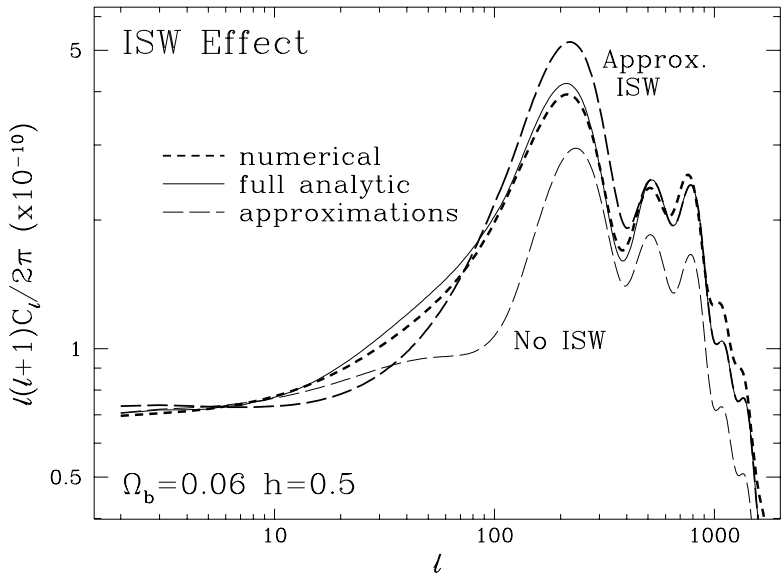

**Figure 7.** The ISW effect. Ignoring the ISW effect entirely leads to a significant error in both the normalization at 10° and shape of the anisotropies due to contributions near recombination. This can be partially accounted for by approximating *all* of the ISW contribution to occur near recombination. This approximation leads to $10 - 15\%$ errors in temperature due to the fact that some of the contribution comes from more recent times where the fluctuation subtends a larger angle angle on the sky. The full integration therefore has more power at larger angular scales and makes the rise to the first Doppler peak more gradual.

In Fig. 7, we show the results of ignoring the ISW integral, using equation (A–23), and integrating the ISW effect with the derivative of equation (A–22). Ignoring the ISW contribution entirely is clearly a bad approximation at the $20 - 25\%$ level in temperature fluctuations since it seriously misestimates the normalization and the shape of the rise toward the first Doppler peak. Employing the approximation improves the situation to the $10 - 15\%$ level in temperature. This approximation has the effect of putting the power at too high a multipole $\ell$ corresponding to the angle that the scale subtends at last scattering rather than the true distance. We see that detailed knowledge of the form of the derivatives only represents a $10 - 15\%$ shift in power which justifies our crude approximation of it.

## Appendix B: The Tight Coupling Approximation

*B.1 The WKB Approximation*

The tight coupling approximation has often been used in the past to describe the behavior of fluctuations at sufficiently small scales that gravitational effects can be neglected entirely (*e.g.* Peebles & Yu 1970). Here we show that it is easy to include the effects of a realistic time dependent potential in the tight coupling formalism even in the intermediate regime where both gravity and pressure play a role. In short, the technique involves expanding the Boltzmann equation (1) and Euler equation (3) in the Compton scattering time $\dot{\tau}^{-1}$ in the limit that Compton scattering is much more rapid than expansion or gravitational infall.

To zeroth order, we regain the tight coupling identities,

$$\Theta_0(\eta) = \Theta(\eta, \mu) + i\mu\Theta_1(\eta),$$
$$\Theta_1(\eta) = V_b(\eta),$$
(B-1)

where we have used the $\ell$th moment of equation (1) to show that $\Theta_\ell = 0$ for $\ell \geq 2$. These equations merely express the fact that the radiation is isotropic in the baryon rest frame. Substituting the zeroth order solutions back into equations (1) and (3), we obtain the iterative first order solution,

$$\dot{\Theta}_0 = -\frac{k}{3}\Theta_1 - \dot{\Phi},$$
$$\dot{\Theta}_1 = -\frac{\dot{R}}{1+R}\Theta_1 + \frac{1}{1+R}k\Theta_0 + k\Psi,$$
(B-2)

where we have used the relation $\dot{R} = (\dot{a}/a)R$. Note that we have used the tight coupling approximation to eliminate the multiple time scales and the infinite hierarchy of coupled equations of the full problem. In fact, this simple set of equations can readily be solved numerically [see *e.g.* Seljak (1994)]. This may be desirable in the case where one needs to know very accurately the transition regime between the large and small scale analytic approximations.

For the purpose of obtaining analytic solutions, it is preferable to rewrite equation (B-2) as a single second order equation,

$$\ddot{\Theta}_0 + \frac{\dot{R}}{1+R}\dot{\Theta}_0 + k^2 c_s^2 \Theta_0 = F(\eta),$$
(B-3)

where

$$F(\eta) = -\ddot{\Phi} - \frac{\dot{R}}{1+R}\dot{\Phi} - \frac{k^2}{3}\Psi,$$
(B-4)

is the forcing function with $\ddot{\Phi}$ as the ISW effect, $\dot{\Phi}$ as the modification to expansion damping, and $\Psi$ as the gravitational infall. The homogeneous $F(\eta) = 0$ equation yields the two fundamental solutions under the WKB approximation,

$$\theta_a(\eta) = (1+R)^{-1/4}\cos kr_s,$$
$$\theta_b(\eta) = (1+R)^{-1/4}\sin kr_s,$$
(B-5)

where the sound horizon $r_s = \int c_s d\eta$ can be integrated to give

$$k_{eq}r_s(\eta) = \frac{2}{3}\sqrt{\frac{6}{R(\eta_{eq})}}\ln\frac{\sqrt{1+R(\eta)}+\sqrt{R(\eta)+R(\eta_{eq})}}{1+\sqrt{R(\eta_{eq})}},$$
(B-6)

with $k_{eq} = \sqrt{2}(\Omega_0 H_0^2 a_0/a_{eq})^{1/2} = (4-2\sqrt{2})/\eta_{eq}$ as the scale that enters the horizon at equality. The phase relation just reflects the nature of acoustic oscillations. If the sound speed were constant, it would yield the expected dispersion relation $\omega = kc_s$.

Now we need to take into account the forcing function $F(\eta)$ due to the gravitational potentials $\Psi$ and $\Phi$. Employing the Green's method, we construct the particular solution,

$$\hat{\Theta}_0(\eta) = C_1\theta_a(\eta) + C_2\theta_b(\eta) + \int_0^\eta \frac{\theta_a(\eta')\theta_b(\eta) - \theta_a(\eta)\theta_b(\eta')}{\theta_a(\eta')\dot{\theta}_b(\eta') - \dot{\theta}_a(\eta')\theta_b(\eta')}F(\eta')d\eta'.$$
(B-7)

Employing equation (B-5) yields

$$\theta_a(\eta')\theta_b(\eta) - \theta_a(\eta)\theta_b(\eta') = [1+R(\eta)]^{-1/4}[1+R(\eta')]^{-1/4}\sin[kr_s(\eta) - kr_s(\eta')],$$
(B-8)

and

$$\theta_a(\eta')\dot{\theta}_b(\eta') - \dot{\theta}_a(\eta')\theta_b(\eta') = \frac{k}{\sqrt{3}}[1+R(\eta')]^{-1}.$$
(B-9)

Equation (8) now follows by fixing $C_1$ and $C_2$ with the initial conditions.

*B.2 Large and Small Scale Corrections to the WKB Approximation*

On large scales the WKB approximation breaks down, whereas on small scales we need to modify equation (B–2) to account for photon diffusion. The WKB approximation assumes that the frequency, *i.e.* the sound speed, is not rapidly varying

$$(kc_s)^2 \gg (1+R)^{1/4}\frac{d^2}{d\eta^2}(1+R)^{-1/4}, \qquad \text{(B-10)}$$

and is valid if $R \ll 1$ and on small scales. Roughly speaking, this requires the mode to be in the oscillatory region by last scattering. More specifically, it is reasonably well satisfied at recombination if $k_s > 0.08h^3$ for the range of $\Omega_b$ consistent with nucleosynthesis. Since we know the solution at scales $k \ll k_s$ where we can neglect pressure, we can obtain the full solution by matching the two. Let us see how this is done.

At large scales, we know the only the ISW effect is important since $\ell$-mode coupling and infall are only effective inside the horizon. The solution to equation (B–2) is therefore

$$\Theta_0(\eta) = \Theta_0(0) - \int_0^\eta \dot{\Phi}(\eta')d\eta'. \qquad \text{(B-11)}$$

On the other hand, the WKB approximation at large scales, *i.e.* the $kr_s \to 0$ limit, predicts

$$(1+R)^{1/4}\hat{\Theta}_0(\eta) = \Theta_0(0) - \int_0^\eta \dot{\Phi}(\eta')\frac{\sqrt{3}}{k}\frac{d}{d\eta'}\left\{[1+R(\eta')]^{3/4}k[r_s(\eta')-r_s(\eta)]\right\}d\eta', \qquad \text{(B-12)}$$

where we have integrated once by parts and employed $\Theta_0(0) = -\Phi(0)$. If $R = 0$, $\dot{r}_s = c_s = 1/\sqrt{3}$ and the two expressions are identical. Therefore, the large scale solution obeys equation (B–7) if the two fundamental solutions are taken to be

$$\begin{aligned}\theta_a(\eta) &= \cos kr_s.\\ \theta_b(\eta) &= \sin kr_s,\end{aligned} \qquad \text{(B-13)}$$

This is a particularly useful form to express the large scale solution because in the CDM model, $R(\eta_*) \ll 1$ since $\Omega_0 \gg \Omega_b$. This solution will therefore approximately join into the WKB approximation at small scales. We take the matching point to be $k_s = 0.08h^3$ but the results are not terribly sensitive to what scale is chosen (*c.f.* Fig. 2). In fact for a simple approximate estimate ($\sim 20\%$ in $\Delta T/T$) to the CDM spectrum, one can employ equations (B–13) on all scales. More explicit formulae can be found in Appendix D.

A correction to the tight coupling equation (B–3) itself must be applied at small scales. Notice that the finite time scale of Thomson scattering, controlling the mean free path of the photons, has not appeared anywhere in the analysis so far. That is because we have only expanded to first order in $\dot{\tau}^{-1}$. For diffusion effects, we need solve iteratively to second order. In this case, we do neglect the effects of the ISW term and gravitational infall for the baryons. The dispersion relation must consequently be modified to have an imaginary term. If we write the solution for $\Theta_0 + \Psi$ as $\exp(i\phi_D)$ we obtain [see *e.g.* Peebles (1980)]

$$\phi_D(\eta) = \pm k\int_0^\eta c_s d\eta + i[k/k_D(\eta)]^2, \qquad \text{(B-14)}$$

where the damping length is given by

$$k_D^{-2}(\eta) = \frac{1}{6}\int d\eta \frac{1}{\dot{\tau}}\frac{R^2 + 4(1+R)/5}{(1+R)^2}. \qquad \text{(B-15)}$$

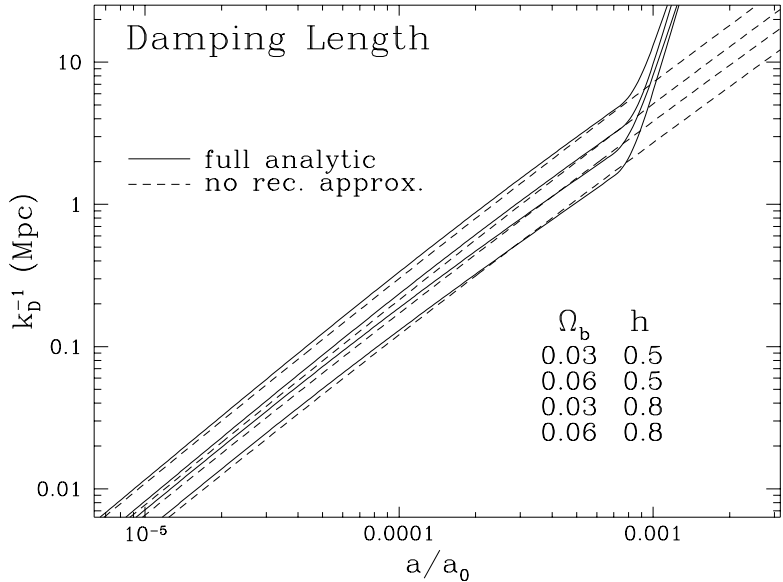

**Figure 8.** The evolution of damping through recombination. The damping length increases rapidly at recombination corresponding to the fact that the mean free path of the photons becomes infinite. Weighting by the visibility function, which tells us when the photon last scattered, yields the average damping factor. We have also included a simple approximation to the damping integral which can be used for estimation purposes *before* recombination.

This then yields the damping correction given in equation (10). Before recombination, this integral may be simply approximated as

$$k_D^{-2} \approx 1.7 \times 10^7 (1 - Y_p/2)^{-1} (\Omega_b h^2)^{-1} (\Omega_0 h^2)^{-1/2} \left(\frac{a}{a_0}\right)^{5/2} \frac{1}{3\sqrt{a_{eq}/a} + 2} \mathrm{Mpc}^2 \qquad a \ll a_*, \qquad \text{(B-16)}$$

where $Y_p \approx 0.23$ is the primordial helium mass fraction. This may be useful for estimation purposes but should *not* be used to describe the detailed damping process at recombination since the damping length suddenly increases to infinity. To illustrate this effect, in Fig. 8 we plot the evolution of the damping scale through recombination obtained by following the true ionization history obtained in Appendix C in comparison with equation (B-16).

## Appendix C: Recombination

Following Peebles (1968) and Jones & Wyse (1985), we solve for the ionization history through recombination. Since we wish to obtain the detailed behavior of photon diffusion damping including its dependence on cosmological parameters, we need to improve upon the fitting formulae obtained by Jones & Wyse (1985). This is especially necessary in CDM scenarios where $\Omega_b \ll \Omega_0$. Note that the full numerical treatment employs the numerical values for the ionization history rather than the approximations presented here.

The total optical depth from the present to the critical recombination epoch $800 < z < 1200$ can be approximated as

$$\tau(z, 0) \approx \Omega_b^{c_1} \left(\frac{z}{1000}\right)^{c_2} \qquad \text{(C-1)}$$

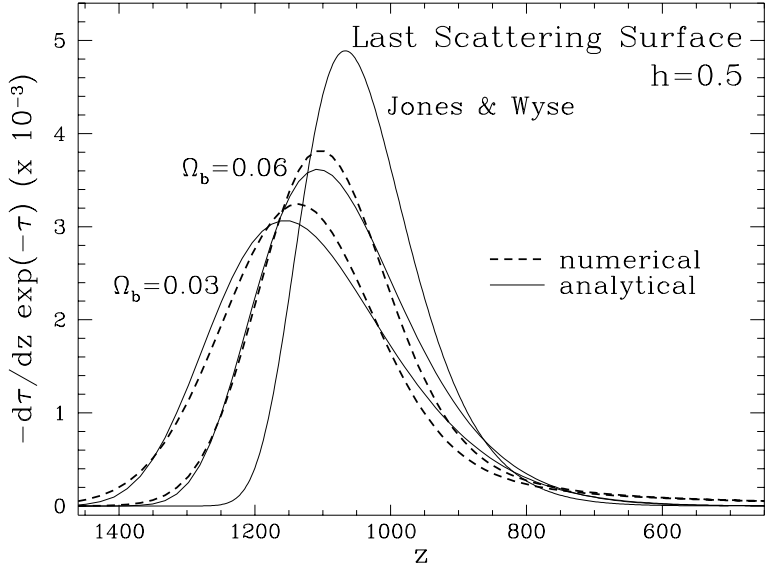

**Figure 9.** The redshift visibility function. Notice that the weak dependence on $\Omega_b$ of the visibility function is adequately described by the analytic fitting formula, whereas the Jones & Wyse (1985) fitting formula [their equation (23)] does not.

where $c_1 = 0.43$ and $c_2 = 16 + 1.8 \ln \Omega_b$. Since the range of reasonable values for $h$ is limited to be $0.5 \lesssim h \lesssim 0.8$, we have ignored the small $h$ dependence. For definiteness, we take last scattering to occur at $z_*$ where the optical depth $\tau(z_*, 0) = 1$. It immediately follows from (C-1) that this occurs at

$$\frac{z_*}{1000} \approx \Omega_b^{-c_1/c_2} = \Omega_b^{-0.027/(1+0.11\ln \Omega_b)} \tag{C-2}$$

which is weakly dependent on $\Omega_b$. The differential optical depth $\dot\tau$ then becomes

$$\dot\tau(z) = \frac{c_2}{1000}\Omega_b^{c_1} \left(\frac{z}{1000}\right)^{c_2-1} \frac{\dot a}{a}(1+z), \tag{C-3}$$

where $\dot\tau$ is by definition positive since $\dot\tau \equiv d[\tau(\eta', \eta)]/d\eta$. Finally the ionization fraction is given by $x_e(z) = \dot\tau a_0 / n_e \sigma_T a$ where

$$(n_e \sigma_T a/a_0)^{-1} = 4.3 \times 10^4 (1 - Y_p/2)^{-1} (\Omega_b h^2)^{-1} (1+z)^{-2} \text{Mpc} \tag{C-4}$$

with $Y_p \approx 0.23$ as the primordial helium mass fraction. Of course, where the formula (C-3) implies $x_e > 1$, set $x_e = 1$, i.e. $\dot\tau = n_e \sigma_T a/a_0$. In Fig. 9, we show the numerical values for the visibility function in redshift space $-(d\tau/dz)e^{-\tau}$ compared with these analytic fits.

The decrease in ionization fraction implies an increase in the distance a photon can diffuse and damp (see Fig. 8). This is evaluated by employing the above formulae for the ionization fraction in equation (10). Recombination therefore causes a sudden increase in the damping of anisotropies. In Fig. 10, we display the effect of recombination on anisotropies by comparing the full solution for the damping with that obtained by assuming instantaneous recombination, i.e. approximating the damping factor as $\mathcal{D}(k) \approx \exp\{-[k/k_D(\eta_*)]^2\}$ where $k_D^{-2}(\eta)$ is given by equation (B-16). The misestimation of the damping scale is significant, but it is still a much better approximation than neglecting damping entirely.

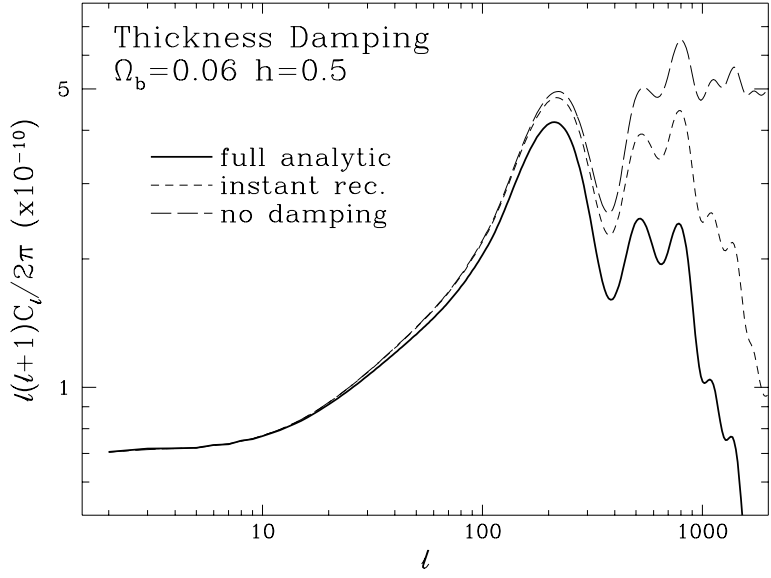

**Figure 10.** The effect of the finite thickness of the last scattering surface. Estimating the damping in the instantaneous recombination approximation leads to a significant underestimate of the damping scale. It is however far better than neglecting diffusion damping entirely.

# Appendix D: A User's Manual

*D.1 Explicit Tight Coupling Expressions*

In this Appendix, we bring together and list the various ingredients necessary for the analytic solution that have been scattered throughout the text and other appendices. First of all, although equation (8) is the best way to write the solutions in the tight coupling limit for understanding the physics of CMB anisotropies, for calculational purposes it is convenient to express the solutions in a more explicit but cumbersome form. One advantage of the analytic tight coupling solutions is they do not require the use of time derivatives of the potentials despite the appearance of equation (8). Thus accuracy is not compromised by our lack of a detailed description for $\dot\Phi$ and $\dot\Psi$. Integrating equation (8) by parts twice, we obtain

$$(1+R)^{1/4}[\hat\Theta_0(\eta) + \Phi(\eta)] = [\cos kr_s(\eta) + J(0)\sin kr_s(\eta)][\Theta_0(0) + \Phi(0)] + I(\eta),\qquad(D\text{-}1)$$

where the overhat denotes the undamped solution,

$$J(\eta) \equiv -(1+R)^{3/4}\frac{\sqrt{3}}{k}\frac{d}{d\eta}(1+R)^{-1/4} = \frac{\sqrt{3}}{4k}\frac{\dot R}{\sqrt{1+R}},\qquad(D\text{-}2)$$

and

$$I(\eta) = \frac{k}{\sqrt{3}}\int_0^\eta d\eta'\,\Phi(\eta')G(\eta')\sin[kr_s(\eta) - kr_s(\eta')],\qquad(D\text{-}3)$$

with

$$G(\eta) = (1+R)^{-1/4}\left[1 - (1+R)\frac{\Psi}{\Phi} + \frac{3}{4k^2}\ddot R - J^2\right].\qquad(D\text{-}4)$$

Here we have employed the identity $\dot{\hat\Theta}_0(0) = -\dot\Phi(0)$. Since the ISW effect predicts constant $\Theta_0 + \Phi$ at superhorizon scales, we have written these expressions in terms of that quantity.

The dipole solution $\hat{\Theta}_1$ can be similarly obtained from the photon continuity equation $k\Theta_1 = -3(\dot{\Theta}_0+\dot{\Phi})$,

$$(1+R)^{3/4}\frac{\hat{\Theta}_1(\eta)}{\sqrt{3}} = [1+J(\eta)J(0)][\Theta_0(0)+\Phi(0)]\sin kr_s(\eta)$$
$$+ [J(\eta) - J(0)][\Theta_0(0)+\Phi(0)]\cos kr_s(\eta) \qquad \text{(D-5)}$$
$$+ J(\eta)I(\eta) - \frac{k}{\sqrt{3}}\int_0^\eta d\eta \Phi(\eta')G(\eta')\cos[kr_s(\eta) - kr_s(\eta')],$$

where we have used the relation $\dot{r}_s = c_s = (1/\sqrt{3})(1+R)^{-1/2}$. Notice that we do not need $\dot{\Phi}$ even in the boundary terms in either equation (D-1) and (D-5). These forms also bring out the fact that whereas the monopole term is $\propto (1+R)^{-1/4}$, the dipole being $\propto (1+R)^{-3/4}$ is suppressed in comparison.

On the other hand, the large scale solution discussed in Appendix B may be obtained by dropping $R$. To be explicit, these are

$$[\hat{\Theta}_0(\eta) + \Phi(\eta)] = [\Theta_0(0)+\Phi(0)]\cos kr_s(\eta) + \frac{k}{\sqrt{3}}\int_0^\eta d\eta'[\Phi(\eta') - \Psi(\eta')]\sin[kr_s(\eta) - kr_s(\eta')], \qquad \text{(D-6)}$$

and

$$\hat{\Theta}_1(\eta)/\sqrt{3} = [\Theta_0(0)+\Phi(0)]\sin kr_s(\eta) - \frac{k}{\sqrt{3}}\int_0^\eta d\eta'[\Phi(\eta') - \Psi(\eta')]\cos[kr_s(\eta) - kr_s(\eta')]. \qquad \text{(D-7)}$$

Finally, the following relations are useful for computation:

$$R = \frac{1}{1-f_\nu}\frac{3}{4}\frac{\Omega_b}{\Omega_0}a, \quad \dot{R} = \dot{a}R(\eta_{eq}) = \frac{k_{eq}}{\sqrt{2}}\sqrt{1+a}R(\eta_{eq}), \quad \ddot{R} = \frac{1}{4}k_{eq}^2 R(\eta_{eq}), \qquad \text{(D-8)}$$

where we have employed the relation $k_{eq}\eta = 2\sqrt{2}(\sqrt{1+a} - 1)$. Here $1 + \rho_\nu/\rho_\gamma = (1-f_\nu)^{-1} = 1.68$. Note that the scale factor is normalized at equality $a_{eq}/a_0 = a_0^{-1} = 4.0 \times 10^{-5}(T_0/2.7\text{K})^4(\Omega_0 h^2)^{-1}$, and the scale which passes the horizon at equality is $k_{eq} = 1.17/\eta_{eq} = 7.46 \times 10^{-2}(T_0/2.7\text{K})^{-2}\Omega_0 h^2$ Mpc$^{-1}$, with $T_0$ as the present temperature of the CMB. We have gathered together many of these commonly used symbols, and the equations in which they are defined or first appear, in Table 1.

### D.2. Poor Man's Boltzmann Code: A Recipe

Now let us outline the steps in the analytic calculation:

[1] For $k > k_s = 0.08h^3$, take the tight coupling solutions for the undamped monopole $\hat{\Theta}_0$ and dipole $\hat{\Theta}_1$ from equations (D-1) and (D-5) with the potentials from equation (A-17) and (A-22) and the initial conditions from equation (A-19) and (A-20). Evaluate this at last scattering $\eta_*$ given by equation (C-2).

[2] For $k < k_s = 0.08h^3$, repeat the steps in [1] using the large scale solutions equations (D-6) and (D-7) in place of equations (D-1) and (D-5). Join the two solutions at $k_s$.

[3] Evaluate the damping scale function $k_D(\eta)$ from equation (10). Use the recombination fitting formula (C-3) for $\dot{\tau}$ where it implies the ionization fraction $x_e < 1$ and equation (C-4) for earlier epochs, i.e. $x_e = 1$ and $\dot{\tau} = n_e \sigma_T a/a_0$. Integrate the damping factor against the visibility function $\dot{\tau}\exp(-\tau)$ from equation (C-1) and (C-3) thus calculating $\mathcal{D}(k)$ from equation (13). This damping factor is defined such that the true anisotropy at last scattering $[\Theta_0 + \Psi](\eta_*) = \mathcal{D}(k)[\hat{\Theta}_0 + \Psi](\eta_*)$ and $\Theta_1(\eta_*) = \mathcal{D}(k)\hat{\Theta}_1(\eta_*)$.

[4] Free stream the perturbation from last scattering to the present by employing equation (12). For an added 10% correction in the low $h$ CDM scenarios, perform the ISW integral in equation (12) using the derivative of the potentials in equation (A-22).

[5] Approximate the integral over $k$ modes in equation (14) as a sum to obtain the total anisotropy. Since the integrand is oscillatory, unless a large number of $k$ values are taken, spurious oscillations will occur in $C_\ell$, both in the analytic and numerical calculations. This is in practice not a problem since weighting from any realistic experimental window function will automatically smooth out these oscillations. Rather than take extra computational time for purely aesthetic reasons, we employ only of order 100 $k$ values between the present horizon and the damping scale at recombination but smooth the final results in $\ell$ for display. One can verify that this is a valid procedure by increasing the number of $k$ divisions.

[6] Normalize $C_\ell$ to the COBE DMR detection at large scales. Here we have used the rms $10°$ value $(\Delta T/T)^2_{10°} = \sum (2\ell+1) C_\ell W_\ell / 4\pi = 1.25 \times 10^{-10}$ (Bennett *et al.* 1994) where the COBE window function is $W_\ell = \exp[-\ell(\ell+1)\sigma^2]$ with $\sigma = 0.0742$ as the gaussian width of the $10°$ beam. For Harrison-Zel'dovich CDM models, one may alternatively normalize to $Q_{rms-PS} = T_0 (5C_2/4\pi)^{1/2} = 20\mu K$ (Gorski, *et al.* 1994).

On the other hand, if only a rough estimate of anisotropies in the CDM model (20% in temperature fluctuations for scales up to the second Doppler peak) is needed, the following quick and easily coded procedure can be used:

[a] As [2] above, evaluate $\hat{\Theta}_0$ and $\hat{\Theta}_1$ in the large scale limit but employ this for the small scales as well. This amounts to a $10-20\%$ discrepancy in the oscillation behavior in a CDM model where $R(\eta_*) \ll 1$. Use the simple damping factor from the instantaneous recombination approximation $\mathcal{D}(k) = \exp\{-[k/k_D(\eta_*)]^2\}$ with $k_D^{-2}(\eta)$ from equation (B–16). See Fig. 10 for the error this causes.

[b] Free stream the solution by approximating it with equation (A–23) (10% $\Delta T/T$ errors, see Fig. 7). Follow steps [5] and [6] above.

**Table 1: Commonly Used Symbols**

| Symbol | Definition | Equation |
|---|---|---|
| $\Delta_T$ | Total density fluctuation | (A-1) |
| $\Theta$ | CMB temperature fluctuation | (1) |
| $\Theta_0$ | CMB Monopole fluctuation | (1) |
| $\Theta_1$ | CMB Dipole fluctuation | (1) |
| $\Theta_\ell$ | CMB $\ell$th multipole fluctuation | (1) |
| $\Pi$ | Anisotropic stress perturbation | (A-3) |
| $\Phi$ | Gravitational (Newtonian) potential | (A-1) |
| $\Psi$ | Gravitational (curvature) potential | (A-2) |
| $\eta$ | Conformal time | (1) |
| $\eta_0$ | Present conformal time | (11) |
| $\eta_*$ | Recombination conformal time | (C-2) |
| $\sigma_T$ | Thomson cross section | (1) |
| $\tau$ | Thomson optical depth | (1) |
| $\mathcal{D}$ | Temperature damping factor | (13) |
| $A$ | Initial power spectrum | (A-20) |
| $C_\ell$ | Anisotropy power spectrum | (14) |
| $F$ | Gravitational driving force | (5) |
| $N_\ell$ | Neutrino $\ell$th multipole | (A-3) |
| $R$ | Normalized scale factor $3\rho_b/4\rho_\gamma$ | (3) |
| $U_D$ | $\Pi = 0$ decaying mode | (A-7) |
| $U_G$ | $\Pi = 0$ growing mode | (A-6) |
| $V_T$ | Total velocity amplitude | (A-4) |
| $V_b$ | Baryon velocity amplitude | (1) |
| $a$ | Scale factor | (1) |
| $a_0$ | Present scale factor | (1) |
| $a_{eq}$ | Equality scale factor | (D-8) |
| $c_s$ | Photon-baryon sound speed | (6) |
| $f_\nu$ | Neutrino fraction $\rho_\nu/(\rho_\nu + \rho_\gamma)$ | (A-13) |
| $k$ | Fourier mode wavenumber | (1) |
| $k_D$ | Diffusion damping wavenumber | (10) |
| $k_{eq}$ | Equality horizon wavenumber | (D-8) |
| $k_s$ | Solution switching wavenumber | (B-10) |
| $\ell$ | Multipole number | (2) |
| $n_e$ | Electron number density | (1) |
| $r_s$ | Sound horizon | (7) |
| $x_e$ | Electron ionization fraction | (1) |
| overbar | Undamped (pressureless) solution | (A-6) |
| overdot | Conformal time derivative | (1) |
| overhat | Undamped (diffusionless) solution | (8) |